\documentclass[article,twocolumn,superscriptaddress,amsfonts,amssymb]{revtex4}

\usepackage{graphicx}
\usepackage{dcolumn}
\usepackage{bm}
\usepackage{times}
\usepackage{amsmath}
\usepackage{float}

\usepackage{color}

\definecolor{drkgr}{rgb}{0.05,0.6,0.2}
\definecolor{drkred}{rgb}{0.80,0.2,0.2}
\definecolor{light-gray}{gray}{0.55}

\newcolumntype{/}{D{/}{/}{2,2}}  
\newcolumntype{.}{D{.}{.}{0}}  

\begin{document}

\title{
%
Metamagnetic texture in a polar antiferromagnet
%
%
}

\author{D. A. Sokolov$^{\ast}$}
\affiliation{Max-Planck-Institut f\"{u}r Chemische Physik fester Stoffe, D-01187 Dresden, Germany}

\author{N. Kikugawa}
\affiliation{National Institute for Materials Science, Tsukuba 305-0003, Japan}

\author{T. Helm}
\affiliation{Max-Planck-Institut f\"{u}r Chemische Physik fester Stoffe, D-01187 Dresden, Germany}

\author{H. Borrmann}
\affiliation{Max-Planck-Institut f\"{u}r Chemische Physik fester Stoffe, D-01187 Dresden, Germany}

\author{U. Burkhardt}
\affiliation{Max-Planck-Institut f\"{u}r Chemische Physik fester Stoffe, D-01187 Dresden, Germany}

\author{R. Cubitt}
\affiliation{Institut Laue-Langevin, 6 Rue Jules Horowitz, F-38042 Grenoble, France}

\author{J. S. White}
\affiliation{Laboratory for Neutron Scattering and Imaging (LNS), Paul Scherrer Institute (PSI), CH-5232 Villigen, Switzerland}

\author{E. Ressouche}
\affiliation{Univ. Grenoble Alpes, CEA, INAC-MEM, 38000 Grenoble, France}

\author{M. Bleuel}
\affiliation{NIST Center for Neutron Research
National Institute of Standards and Technology
Gaithersburg, MD 20988-8562, USA}
\affiliation{Department of Materials Science and Engineering
University of Maryland, College Park, MD  20742-2115, USA}

\author{K. Kummer}
\affiliation{ESRF, 71 avenue des Martyrs, 38000 Grenoble, France}

\author{A. P. Mackenzie}
\affiliation{Max-Planck-Institut f\"{u}r Chemische Physik fester Stoffe, D-01187 Dresden, Germany}
\affiliation{Scottish Universities Physics Alliance (SUPA), School of Physics and Astronomy, University of St Andrews, St Andrews KY16 9SS, United Kingdom}

\author{U. K. R\"o\ss ler}
\affiliation{IFW Dresden, PO Box 270116, D-01171 Dresden, Germany}

\begin{abstract}
The notion of a simple ordered state implies homogeneity.
If the order is established by a broken symmetry, the elementary Landau theory
of phase transitions shows that only one symmetry mode describes this state.
Precisely at points of phase coexistence domain states
formed of large regions of different phases
can be stabilized by long range interactions.
In uniaxial antiferromagnets the so-called metamagnetism
is an example of such a behavior, when antiferromagnetic
and field-induced spin-polarized paramagnetic/ferromagnetic states
co-exist at a jump-like transition in the magnetic phase diagram.
%
%
Here, combining experiment with theoretical analysis,
we show that a different type of mixed state
between antiferromagnetism and ferromagnetism
can be created in certain non-centrosymmetric materials.
In the small-angle neutron scattering experiments we observe a field-driven spin-state in the layered antiferromagnet Ca$_3$Ru$_2$O$_7$,
which is modulated on a scale between 8 and 20 nm and has both antiferromagnetic and ferromagnetic parts.
We call this state a {\textit{metamagnetic texture}}
and explain its appearance by the chiral twisting effects
of the asymmetric Dzyaloshinskii-Moriya (DM) exchange.
%
%
The observation can be understood as an extraordinary coexistence,
in one thermodynamic state, of spin-orders
belonging to different symmetries.
Experimentally, the complex nature of this metamagnetic state is demonstrated by measurements of anomalies
in electronic transport which reflect the spin-polarization in the metamagnetic texture, determination of the magnetic orbital moments, which supports
the existence of strong spin-orbit effects, a pre-requisite for the mechanism of twisted magnetic states in this material.
Our findings provide an example of a rich and largely unexplored
class of textured states. Such textures mediate
between different ordering modes near phase co-existence,
and engender extremely rich phase diagrams.
\end{abstract}

\maketitle

\section{Introduction}
The term \textit{metamagnetism}, possibly coined by Kramers as a joke,
was used to describe the bizarre properties
of certain magnetic materials that have been investigated for more than 100 years.
They appeared to be paramagnetic or antiferromagnetic in the ground
state, but ferromagnetic in applied fields\cite{Becquerel1939a,Becquerel1939b,Stryjewski1977}.
\textit{Metamagnetism} now labels a sudden rise or cross-over
of the magnetization under applied field and is observed
in various classes of materials.
Once N\'{e}el's notion of antiferromagnetism had been accepted,
one type of metamagnetic behavior could easily be explained
as the jump-like transition between a collinear antiferromagnetic
up-down state and a spin-polarized up-up state when a field overcomes
the exchange between sublattice spins constrained to collinear configurations
by a strong easy-axis magnetic anisotropy \cite{Neel1957,Stryjewski1977}.
In the generic magnetic phase diagrams of such materials,
a first-order phase transition occurs between the two spin-orders.
Long-range classical dipolar interactions or magnetostrictive
interactions can stabilize domain states in which the two spin-orders co-exist.
These classical domain structures at phase-coexistence points
are well understood \cite{Baryakthar1988}.

Ordered states with rotatable order parameters may also display intrinsically
inhomogeneous phases. Here twisting short-range forces cause a continuous modification of the order-parameter direction.
In non-centrosymmetric magnetically ordered materials, spin-orbit effects on the magnetic exchange interactions
cause chiral spiral ordering\cite{Dzyaloshinskii1964a}. Phenomenological theory is able to predict and describe such modulated states in a wide
range of condensed matter systems, such as incommensurable states in certain crystals undergoing
lattice instabilities \cite{Levanyuk1986,Cummins1990},
or chiral liquid crystals \cite{DeGennes1971,Meiboom1981,WrightMermin1989}.
In such modulated textures
the direction of a multicomponent order parameter
spatially rotates from one orientation to another.
Chiral helimagnetic order is a paragon of such textures in which spin-orbit coupling twists an elementary spin-ordered pattern
over long periods \cite{Dzyaloshinskii1964a,BakJensen1980,Nakanishi1980}.
For this type of directional order in systems with a twisting short-range force,
static multidimensional solitons are theoretically predicted\cite{Bogdanov1995},
which now are called chiral skyrmions in the case of
chiral ferromagnets or N\'{e}el antiferromagnets \cite{Bogdanov2002}.
The condensation of such particle-like states
can yield rich phase diagrams \cite{Bogdanov1989,Roessler2006},
which have become a major topic in condensed matter magnetism over the last decade \cite{Muhlbauer2009,Yu2010}.
Although a chiral helimagnet macroscopically behaves
as an antiferromagnet, the primary magnetic order is
a simple ferromagnetic spin order, twisted into a helix over
long distances \cite{BakJensen1980,Nakanishi1980}.

Here we show how a spiral magnetic order emerges in a material with antiferromagnetic order parameter. The spiral propagates in a direction perpendicular to the wavevector of antiferromagnetic order. Materials displaying such complex textures may also host new types of antiferromagnetic skyrmions, which are a subject of intense theoretical and experimental research\cite{Jungwirth2018}. We present the first experimental realization of a magnetic texture composed of an antiferromagnetic ground state and a ferromagnetic spin-polarized state. We identified the layered orthorhombic antiferromagnetic oxide Ca$_3$Ru$_2$O$_7$, as suitable for a focused search for a metamagnetic texture.
This material crystallizes in the non-centrosymmetric polar structure described
by space-group $\textit{Bb2$_1$m}$, which belongs to polar point-group $C_{2v}$. The crystal structure consists of RuO$_{2}$ bilayers with corner sharing RuO$_{6}$ octahedra, which are rotated around the crystallographic $\textit{c}$-axis and tilted with respect to the $\textit{ab}$-plane\cite{yoshida05}.
The basic antiferromagnetic
order-parameter in Ca$_3$Ru$_2$O$_7$ was identified in detailed neutron diffraction studies\cite{wbao08}.
This magnetic-order parameter is described by a simple collinear ordering-mode,
which does not allow for a canting of moments
into a weak ferromagnetic state.
Thus, Ca$_3$Ru$_2$O$_7$ meets the elementary symmetry conditions for modulated magnetism.
A further requirement is relevant spin-orbit couplings, that affect the primary
magnetic order. We have used X-ray magnetic circular dichroism (XMCD) spectroscopy on the Ru ions
to measure its orbital magnetic moment. We find relatively large moments with a ratio of orbital to
spin moment of about 0.15, see SFIG.1 in Supplementary materials in agreement with earlier \cite{Liu2011} and our own theoretical
investigations. This indicates that possibly strong antisymmetric DM
exchange interactions do affect the magnetic order in Ca$_3$Ru$_2$O$_7$.
The material orders antiferromagnetically below the N\'{e}el temperature,
$\textit{T}$$_{N}$ = 56 K with the ordered moments along the $\textit{a}$-axis and the magnetic propagation vector along the $[001]$ direction. Within the bilayer the Ru moments are coupled ferromagnetically, whereas the coupling between the adjacent bilayers is antiferromagnetic \cite{yoshida05,wbao08}. This state is normally referred to as AFM-a. On cooling below 48 K, the ordered moments within the bilayer spontaneously re-orient to point along the $\textit{b}$-axis, and this state is known as AFM-b. The coupling between the adjacent bilayers remains antiferromagnetic. The moment re-orientation is accompanied by the first order structural transition at 48 K\cite{yoshida04}.

\begin{figure*}[ht]
\includegraphics[scale=0.50]{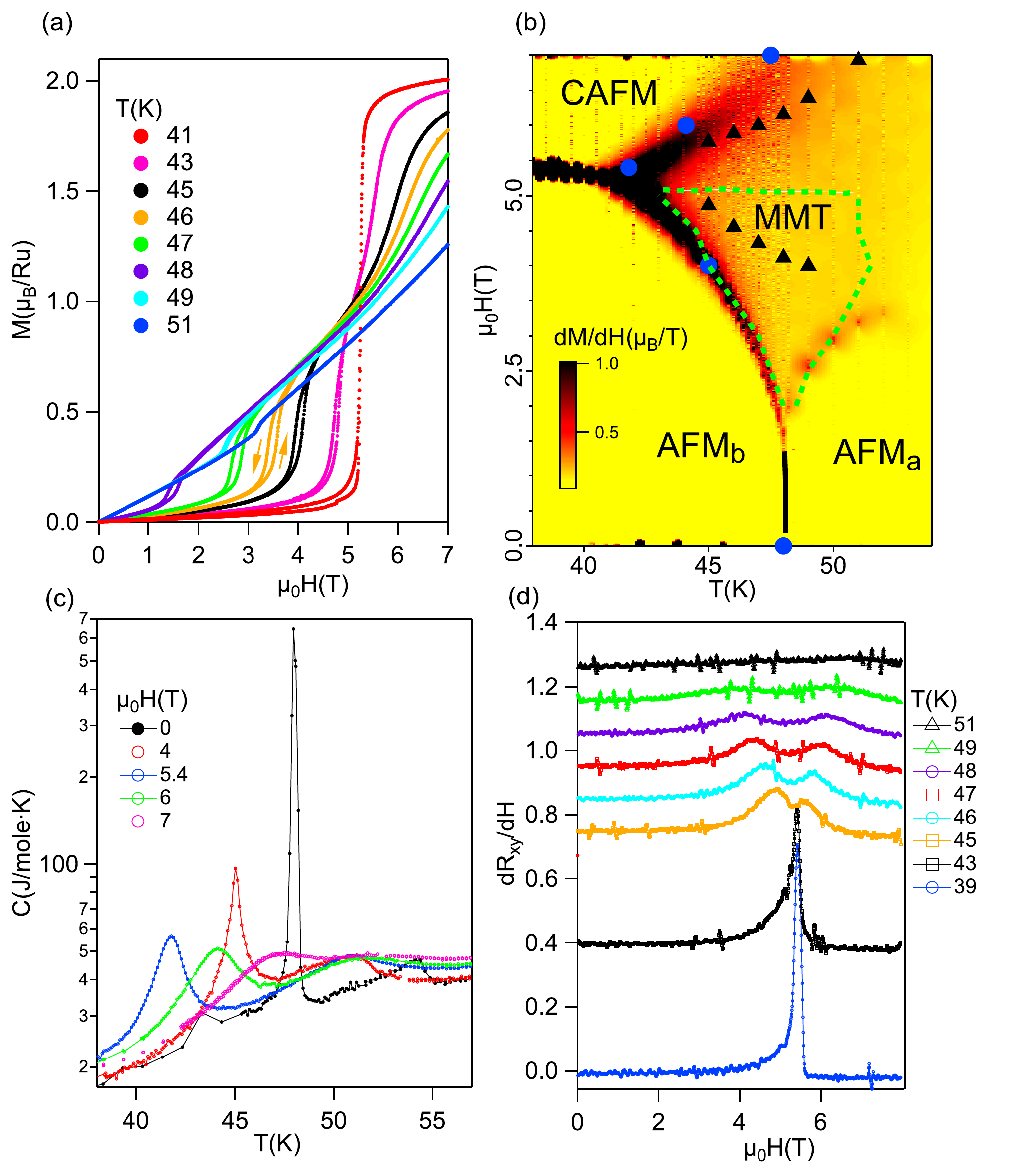}
\caption{
Bulk properties of Ca$_{3}$Ru$_{2}$O$_{7}$ measured with the magnetic field along the b-axis on the same single crystal: (a) Magnetic field dependent magnetization at various temperatures spanning the region of interest in the phase diagram. Up(down) arrow corresponds to increase (decrease) of the field. (b) Differential susceptibility dm/dH obtained by differentiating magnetization in (a) with respect to the field. The color scale represents a magnitude of dm/dH. Dotted green line encircles the region in which the metamagnetic texture (MMT) was observed in SANS measurements, see text. Solid black line corresponds to transitions inferred from low field m(T) measurements. Black filled triangles in (b) correspond to the maximum in dR$_{xy}$/dH in the Hall effect measurements. Blue filled dots in (b) are inferred from the field dependent maxima in the specific heat. AFM-a, AFM-b, and CAFM mark two antiferromagnetic and the canted antiferromagnetic regions of the phase diagram, see text. (c) Temperature dependence of the specific heat. The sharp maximum at zero field marks the moment re-orientation transition. (d) Derivative of the Hall resistivity with respect to the magnetic field.}
\label{bulk}
\end{figure*}

The isothermal magnetization at low temperatures displays a single metamagnetic transition and reaches $\sim$1.95 $\mu$$_{B}$ per Ru ion, a slightly reduced value compared to 2 $\mu$$_{B}$ for the full moment expected for Ru$^{4+}$, FIG~\ref{bulk}a. At T$\geq$43 K the magnetization shows two metamagnetic transitions, which become increasingly separated in field as the temperature increased to 48 K, resulting in the ``funnel''-type structure in the magnetic susceptibility, dm/dH plotted as a function of temperature and magnetic field, FIG~\ref{bulk}b. The higher field transition at H$>$5.5 T in the magnetization is a transition into a canted antiferromagnetic structure, CAFM, in which the magnetic moments on Ru ions are partially polarized along the direction of the applied field\cite{wbao08}. We have also detected a small hysteresis at the low field transition, a typical signature of metamagnetism. Further below we show that the fields and temperatures at which the hysteresis was observed mark the onset of metamagnetic texture observed in small-angle neutron scattering (SANS) experiments.

The metamagnetic transitions between the antiferromagnetic ground states with mutually orthogonal directions of the staggered magnetization have strong signatures in the specific heat measurements, FIG~\ref{bulk}c. A sharp, lambda-like transition in the specific heat in zero field at 48 K is suppressed to lower temperature as the magnetic field increases up to 5.4 T. At higher fields the transition becomes broader and is shifted to higher temperatures in agreement with the ``funnel''-type structure, shown in FIG~\ref{bulk}b.

The metamagnetic transitions also manifest in the electrical transport measurements, as shown in FIG~\ref{bulk}d.
The  Hall resistivity $R_{xy}$ for current along the $\textit{a}$-axis and with magnetic field along the $\textit{b}$-axis
displays two features that are marked by local maxima  in the derivative  $dR_{xy}$/dH.
These maxima split towards lower and higher fields upon increasing temperature,
a behavior similar to dm/dH, see further discussion and SFIG.2,3 in Supplementary materials.
Our analysis, following the general approach for
the Anomalous Hall effect (AHE)\cite{nagaosa2010}, indicates
that in addition to a strong AHE
component there is an intrinsic additional contribution in the region between
the two metamagnetic transitions,
see Supplementary materials.
This suggests the presence of a magnetic texture with either topological
features or non-collinear complex modulations in this magnetic state that
can contribute an extraordinary off-diagonal components of the resistivity tensor.
A quantitative extraction of this extraordinary Hall-resistivity
may only become possible by taking into account field-induced changes
in the band structure and the exact structure of the new magnetic order,
and lies beyond the scope of this work.

The results of our bulk measurements refine the published phase diagram of Ca$_{3}$Ru$_{2}$O$_{7}$ in the region between the lines separating AFM-b and CAFM states\cite{mccall03,gcao04,fobes11}. Until now most of the neutron scattering measurements on Ca$_{3}$Ru$_{2}$O$_{7}$ were performed at commensurate wavevectors. In this Article we focus on the nature of the magnetic state near AFM-a to AFM-b transition at magnetic fields between 2 T and 5 T and report the magnetic modulation in a previously unexplored region of the reciprocal space near the wavevector Q=0.

\begin{figure*}[ht]
\includegraphics[scale=1.0]{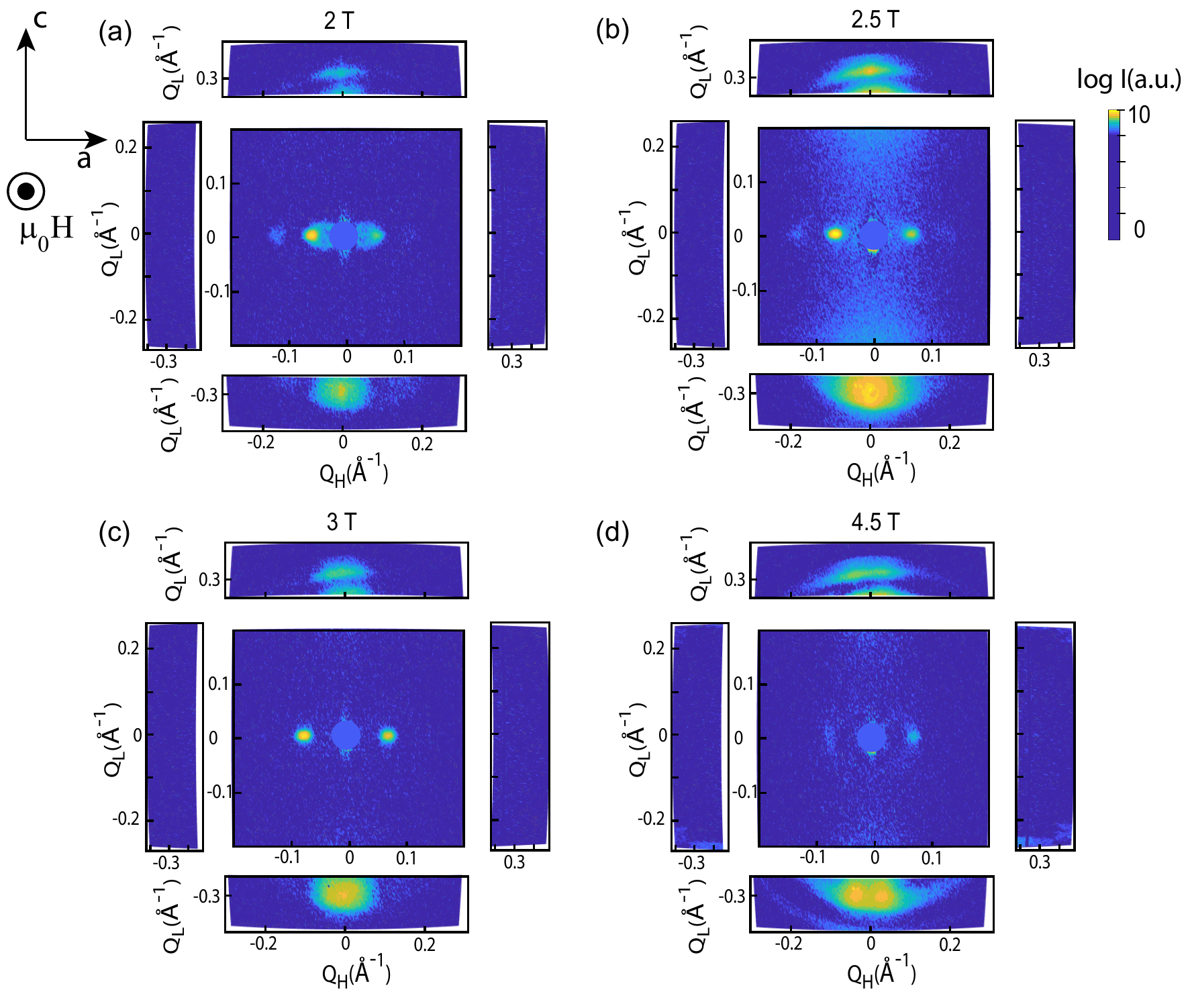}
\caption{
Typical SANS patterns measured at 48 K in magnetic fields from 2 T (a) to 4.5 T (d) applied parallel to the b-axis. 4 smaller panels are SANS detectors positioned at 1.2 m from the sample and thus able to detect diffraction from 001 magnetic reflection. The panel in the centre is the main SANS detector at 2 m from the sample. The metamagnetic texture (MMT) propagates along the $\textit{a}$-axis with Q$_{MMT}$ reaching (0.08,0,0) {\r{A}}$^{-1}$. Higher order reflections were observed at 2 T (a) and 2.5 T (b). A split along the $\textit{a}$-axis reflection at Q$_{AF}$=(0,0,-0.32) {\r{A}}$^{-1}$ was observed at 4.5 T(d) and also at 5 T, see Supplementary materials. No such splitting was observed below the onset of the metamagnetic texture, T$_{MMT}$. Note the same logarithmic scale of the intensity for all fields. The non-magnetic background at 65 K measured at the corresponding field was subtracted from all the patterns. The filled blue circle at Q = 0 is a mask applied to cover the direct neutron beam. Q$_{L}$ and Q$_{H}$ are wavevectors along (00L) and (H00) directions.}
\label{sans}
\end{figure*}

SANS experiments were performed to search directly for a bulk long-wavelength magnetic modulation in the ``funnel''-type region of the $\mu_{0}$H-T phase diagram near Q=0. Typical SANS patterns obtained at 4 different fields are shown in FIG~\ref{sans}. Each pattern consists of the images of the main two-dimensional low-Q detector and 4 additional high-Q detectors as detailed in Ref.\cite{dewhurst2015}. The magnetic field was applied parallel to the $\textit{b}$-axis, which points out of the plane of the detector. Data were collected after zero field cooling the sample to 2 K, applying the field at 2 K and measuring the pattern at several increasing temperatures. For each field a non-magnetic background collected at 65 K was subtracted.
The most striking feature found in our experiments is a pair of satellites at Q$_{MMT}$=($\pm$$\Delta$,0,0), which correspond to a magnetic modulation propagating along the a-axis with a repeat distance of 2$\pi$/$\Delta$, FIG~\ref{sans}. We observed the satellites at fields from 2 T up to 5 T in the temperature range, which shows a hysteretic behaviour of the magnetisation shown in FIG~\ref{bulk}a. No satellites were observed at fields above 5 T, suggesting that the ``funnel''- type region of the phase diagram is not a uniform magnetically ordered state.
The satellites develop from the strong intensity near Q=0 at the temperature of AFM-a to AFM-b transition. The scattering is broad with respect to wavevector near the onset temperature, FIG~\ref{sans}a. The apparent diffuse nature of the scattering is most likely due to a quasi-long-range ordering. The wavevector of satellites initially increases on heating, although in FIG~\ref{sans_ana} we show that the temperature dependence of the wavevector is not monotonic at all fields. This pattern was observed at all fields except for 2 T, where satellites exist only in a very small temperature range in a proximity of the metamagnetic transition. We also observed a second harmonic of the primary satellites at 2Q$_{MMT}$ at 2 T, 2.5 T and 3 T, which could correspond to higher order peaks or represent a double scattering.

The magnetic field and the wavevector dependence of SANS intensity is summarized in FIG~\ref{sans_ana}. Increasing the magnetic field suppresses intensity of satellites at the corresponding temperatures, FIG~\ref{sans_ana}a. The intensity of the satellites is the strongest at the lowest temperature at which we can resolve the satellites from the strong scattering near the direct beam. The wavevector of the modulation is strongly temperature dependent. For $\mu_{0}$H$\geq$4 T it increases continuously with temperature from Q$_{MMT}$$<$0.045 {\r{A}}$^{-1}$ at the onset temperature up to Q$_{MMT}$=0.08 {\r{A}}$^{-1}$ at T$\geq$ 50 K. In contrast, for $\mu_{0}$H$\leq$3 T Q$_{MMT}$ displays a sharp maximum just above the onset temperature, FIG~\ref{sans_ana}b. The repeat distance of the modulation, $\Delta$=2$\pi$/Q$_{MMT}$ reaches $\sim$200 {\r{A}} at the lowest temperature of observation of the satellites at 5 T. Near 50 K, at 5 T the repeat distance decreases to $\sim$80 {\r{A}}. The competing character of several coupling terms, which have different temperature dependencies, is most likely the origin of non-monotonic temperature dependence of the wavevector of the metamagnetic texture, see Supplemental materials for details.
Using the rocking curve measurements we estimated the correlation length of the modulation along the $\textit{b}$-axis, FIG~\ref{sans_ana}c. For details of estimate, see SFIG.4 in supplementary materials. The correlation length $\xi_{b}$ is not resolution limited and reaches 2280 {\r{A}} at 2.5 T, comparable to the correlation length of 5500 {\r{A}} in the A-phase of MnSi\cite{Muhlbauer2009}. Magnetic field suppresses correlations along $\textit{b}$ axis at $\mu_{0}$H$>$2.5 T.
Neutron scattering is only sensitive to the component of the magnetization perpendicular to the total momentum transfer, so that the component of the ordered moment along the $\textit{a}$-axis is not measured in our experiment. The observation of the satellites along the $\textit{a}$-axis in the $\textit{ac}$-plane indicates that the ordered moment of the modulation can have components parallel to the $\textit{c}$ and $\textit{b}$-axes. In AFM-b and AFM-a regions of the phase diagram, which border the region of the metamagnetic texture the ordered moment has no component along the $\textit{c}$-axis. We also note that in none of the Fe and Mn-doped Ca$_{3}$Ru$_{2}$O$_{7}$ does the ordered moment acquire a component along the $\textit{c}$-axis\cite{ke14,ke17}. It is therefore likely that the ordered moment of the modulation is along the $\textit{b}$-axis, but the component along the $\textit{a}$-axis cannot be ruled out. Our modulation is then either a helix or a cycloid if it has a component of the ordered moment along the propagation vector. Further experiments with polarized neutrons are required to identify the type of the modulation.
The strong intensity near Q$_{AF}$=(0,0,$\pm$0.32) {\r{A}}$^{-1}$ or simply (001) corresponds to the bulk antiferromagnetism, which propagates along the $\textit{c}$-axis in agreement with Ref.\cite{wbao08}. The scattering is very broad near Q$_{AF}$ at the temperatures at which the satellites at Q$_{MMT}$ are observed, but turns into a well-defined sharp reflection at the lowest temperatures, where no satellites are observed. The broad, ring-like shaped features near Q$_{AF}$ possibly originate from a short-range or fluctuating antiferromagnetic order. We observed reflection at Q$_{AF}$=(00-1) only, no reflection was observed at Q$_{AF}$=(001) due to a small tilt of the crystal with respect to the vertical direction. We also note that the antiferromagnetic reflection acquires a modulation, which propagates along the $\textit{a}$-axis, which becomes resolvable above 4 T. The wavevector of the modulation, Q$^{m}_{AF}$=($\delta$0-1), where $\delta$ reaches 0.059 {\r{A}}$^{-1}$ at 5 T. This behaviour is reminiscent of a magnetic field-induced commensurate-to-incommensurate transition in the DM antiferromagnet Ba$_{2}$CuGe$_{2}$O$_{7}$, in which the magnetic field was applied in the plane of the rotation of the spins\cite{zheludev98}. A commensurate to incommensurate antiferromagnetic transition was also reported for Fe and Mn-doped Ca$_{3}$Ru$_{2}$O$_{7}$ in Ref.\cite{ke14,ke17}. A cycloidal modulation propagating along the $\textit{a}$-axis was identified for both types of doping. These observations suggest that the antiferromagnetism in Ca$_{3}$Ru$_{2}$O$_{7}$ can be easily destabilized by application of the magnetic field or doping and is prone to host magnetic solitons.

The onset of the magnetic texture with the ordered moment along $\textit{b}$-axis requires the ordered moment in the AFM$_{a}$ state to rotate from the $\textit{a}$-axis to the $\textit{b}$-axis locally, on the lengthscale of the magnetic texture. We propose that such defects in the magnetic structure break the long-range three-dimensional antiferromagnetism in Ca$_{3}$Ru$_{2}$O$_{7}$. The intensity near the antiferromagnetic wavevectors Q$_{AF}$ is maximised near the temperature at which the magnetic satellites at {Q$_{MMT}$ disappear. We note that in a previous neutron scattering work the intensity of the antiferromagnetic reflections measured between 3 and 4 T showed a reduced intensity on cooling \cite{wbao08}. We argue that the emergence of the magnetic texture is the origin of the reduced integrated intensity of the antiferromagnetic Bragg peaks at Q$_{AF}$. It is unlikely therefore, that the magnetic texture and the bulk antiferromagnetism co-exist in a non-equilibrium state. Instead, the magnetic texture develops from the antiferromagnetism as an equilibrium state in the presence of the magnetic field, which enhances the effect of the DM interactions. Further neutron diffraction experiments are needed to describe the splitting of (001) magnetic reflection.

\begin{figure*}[ht]
\includegraphics[scale=0.8]{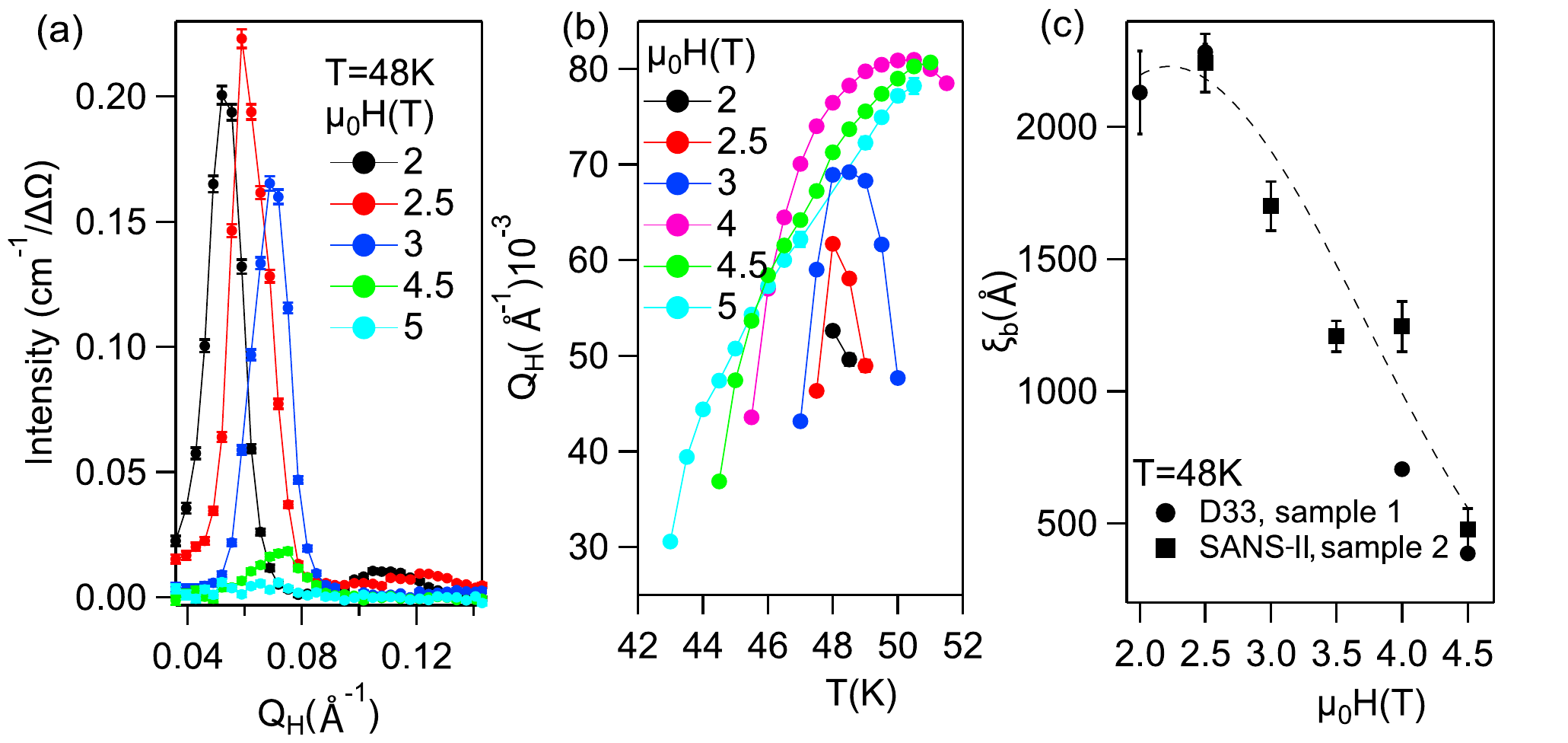}
\caption{%
Metamagnetic textures in Ca$_{3}$Ru$_{2}$O$_{7}$: (a) Magnetic field dependence of the wavevector-dependent azimuthally averaged SANS intensity on the detector plane capturing magnetic satellite at Q$_{MMT}$ measured at 48 K. (b) Temperature dependence of the wavevector of the modulation measured at fields from 2 T to 5 T. (c) Magnetic field dependence of the correlation length along $\textit{b}$-axis at 48 K measured in samples 1 and 2 on D33, ILL and on SANS-II, PSI, see Methods. Where not shown explicitly, the errorbars (one sigma standard deviation) are smaller than markers. Lines are a guide to the eye.}
\label{sans_ana}
\end{figure*}

Summarizing the experiments, Ca$_{3}$Ru$_{2}$O$_{7}$ displays a spirally-modulated magnetic order in broad temperature-field region, previously regarded as a crossover\cite{wbao08,fobes11}. The propagation vector of the spiral is aligned perpendicular to the magnetic field and the staggered magnetization is likely parallel to it. The magnetic field applied along the polar $\textit{b}$-axis destabilizes the antiferromagnetic ground state by flipping spins from the $\textit{a}$-axis to $\textit{b}$-axis locally on a scale between 8 and 20 nm depending on the magnetic field and temperature as illustrated schematically in FIG~\ref{PD4AFMs}b. Then, a mixed state modulating between antiferromagnetic and ferromagnetic spin-configurations with long periods, is observed.
Our observation of metamagnetic textures in Ca$_{3}$Ru$_{2}$O$_{7}$ invites a comparison with spiral spin states driven by a competition between direct exchange interactions such as in Ca$_{3}$Co$_{2}$O$_{6}$\cite{agrestini2008} and MnSc$_{2}$S$_{4}$\cite{gao2017} or rare-earth elements such as Tb, Dy, Ho, which demonstrate a helically modulated magnetic structure due to nesting of the Fermi surface or a Kohn anomaly\cite{jensen1991}. Although the theoretical description of the helical modulation in Ca$_{3}$Co$_{2}$O$_{6}$ is still lacking, it is considered to result from a competition of antiferromagnetic and ferromagnetic direct exchange interactions. A magnetic vortex state reported in MnSc$_{2}$S$_{4}$ results most likely due to a competition between nearest and the next nearest neighbour exchange interactions\cite{gao2017}. Elemental rare-earths such as Tb, Dy, and Ho order via long-range exchange interactions carried by conduction electrons (RKKY) at a finite wavevector dictated by RKKY. As the temperature is lowered the effects of crystalline electric field and magnetic anisotropy lead to a reduction of the ordering wavevector and transition into ferromagnetic state. The theories explaining the magnetic structures of Tb, Dy, Ho consider a high density of states near the Fermi level, which drives the nesting. The phenomenon of metamagnetic textures in Ca$_{3}$Ru$_{2}$O$_{7}$ is distinctly different from both above mentioned examples as the textured state results from a coupling of ferromagnetic and antiferromagnetic order parameters via so-called Lifshitz invariants, see Supplementary materials for details. A rather generic character of such a term in the free-energy expansion suggests that more spin textured states driven by DM interaction are awaiting discovery.
\begin{figure*}[ht]
\includegraphics[scale=0.420]{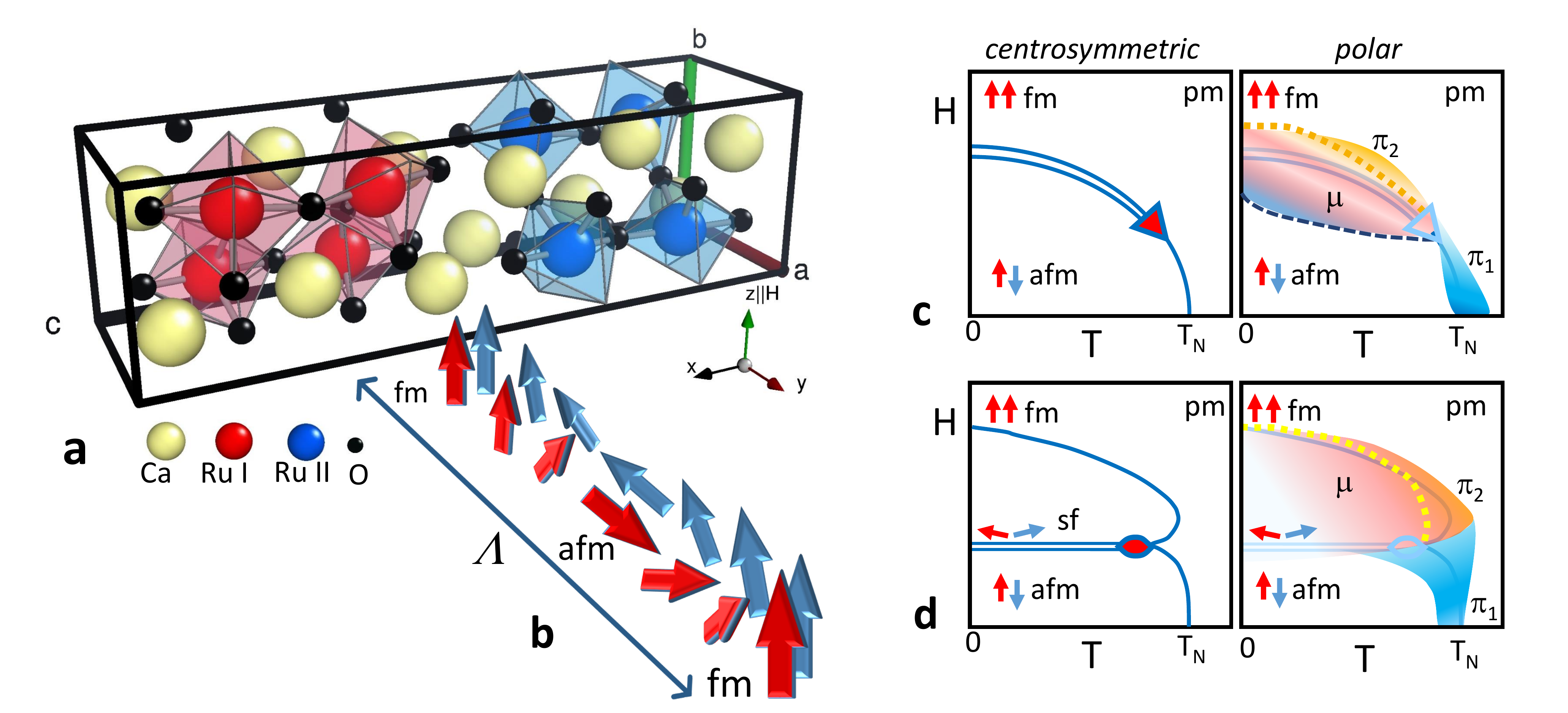}
\caption{%
Crystal structure of Ca$_3$Ru$_2$O$_7$,
panel $\textbf{a}$ shows the
The Ru atoms occupy a single crystallographic site. The antiferromagnetic primary order splits these positions
into two double layers, Ru (I) and Ru (II), which have internally ferromagnetic spin-configurations and are antiparallel.
Above 48 K transition the collinear spin-structure has moments directed along the $\textit{a}$-axis of the orthorhombic cell. The polar axis is $\textit{b}$. Cartesian coordinates $xyz$ are used for spin and spatial gradients, with $z$ along $\textit{b}$ as indicated. Panel $\textbf{b}$ shows a metamagnetic one-dimensional texture propagating along the $\textit{a}$ axis.
In an applied field along the polar axis $\textit{b}$, the spin configuration oscillates
from fm to afm and back to fm over a repeat period $\Lambda$.
$\textbf{c}$, $\textbf{d}$,
schematic phase diagrams of bipartite antiferromagnets with
easy-axis anisotropy.
Panel $\textbf{c}$
the temperature-field phase diagram for the case of
large anisotropy. A first-order transition between an antiferromagnetic (afm) collinear and a spin polarized (fm) state
occurs along the double line. For temperatures above a tricritical point (triangle) the transition
is continuous. The afm-fm co-existence can be replaced in a polar magnet by a region of
modulated phases $\mu$. Towards the paramagnetic state at elevated temperatures, anomalous
transitions into precursor states ($\pi_1, \pi_2$) then are expected.
The dotted yellow line indicates the transition between the
improper, metamagnetic texture and the precursor state $\mu \rightarrow \pi_2$.
The precursor of type
$\pi_1$ only implies modulations of primary afm modes,
while $\pi_2$ states can be metamagnetic,
being composed of modulations between afm and fm modes.
Panel $\textbf{d}$ displays the phase diagram of a system with a weak anisotropy. The double line signifies
a first order transition of the spin-flop type, which is only a re-orientation of the ordered moment.
The marked point is a bi-critical point (almond mark). In an antiferromagnet with a polar structure, metamagnetic
textures $\mu$ can still occur at elevated temperatures and for fields higher than the spin-flop field, when
a sizeable net moment and antiferromagnetic order compete. As in the case of large anisotropy, different types
of precursors, i.e. a proper antiferromagnetic texture $\pi_1$ or metamagnetic textures with coupled
modulations between afm and fm mode $\pi_2$ can occur.
}
\label{PD4AFMs}
\end{figure*}

Elementary considerations are sufficient to
explain why this modulated state exists, and in fact is an expected behavior
in a polar antiferromagnet such as Ca$_{3}$Ru$_{2}$O$_{7}$ when, under a magnetic field,
its magnetic states is transformed into a ferromagnetic configuration.
Its non-centrosymmetric crystal structure and layered antiferromagnetic ordering
(FIG~\ref{PD4AFMs}a)  enable specific couplings
between the two co-existing order parameters,
the antiferromagnetic staggered magnetization and the ferromagnetic spin polarization.
Ultimately, these couplings derive from the DM interactions (DMI) in this material.
The hierarchy between (i) the strong spin-exchange, stabilizing a
certain antiferromagnetic spin-pattern as the ground state, (ii) the twisting
effects of the DMIs on this ground-state, and (iii) the possibility
to tune the system into the spin-polarized state by external magnetic fields,
while temperature is used to tune the weaker magnetic anisotropies,
makes this layered chiral magnet with a polar structure ideally suited
for a modulation of the desired type.

In contrast to a ``proper'' Dzyaloshinskii texture where one directional ordering mode is twisted,
this is a more complex texture generated near the co-existence points of two phases.
Here, the magnetic order is spatially wavering between the co-existing ferromagnetic
and antiferrromagnetic configurations, as sketched in FIG~\ref{PD4AFMs}b.
Therefore,  the term ``metamagnetic texture'' appears
appropriate for this modulated phase in Ca$_{3}$Ru$_{2}$O$_{7}$.

Qualitatively, the mechanism enabling such complex mixed states
can be stated by using symmetries to construct the phenomenological
continuums theory for these ordering modes, i.e. by constructing the Landau-Ginzburg
free energy for the coexisting and coupled ordering modes as reported for Ca$_{3}$Ru$_{2}$O$_{7}$ in the Supplementary materials.
The specific mechanism then is described by free energy terms known as ``Lifshitz-type invariants''.
Such terms are linear in spatial gradients of one mode and couple it to the other mode.
These Lifshitz-type terms describe a frustrated coupling between different pure modes.
The expectation that such terms cause modulations of thermodynamic mixed phases
has been put forth theoretically for a long time
\cite{Levanyuk1976,Stefanovskii1986,Aizu1989,Yablonskii1990,Zavorotnev2002,Milward2005}.
However, these effects can become relevant only if a system can be tuned towards
special multicritical regions of the phase diagram, where the two primary modes co-exist.
This may be the reason why concrete examples for the effects of such terms have been scarce.
Typically such effects have been discussed for
frustrated (magnetic) systems of low symmetry where different order parameters already
co-exist in the ground state\cite{Baryakthar1985}.
Recently, the importance of such couplings has been
raised in the context of ``phase co-existence'' in materials
with multiple electronic instabilities
such as the manganite or cuprate perovskites\cite{Milward2005}.
The phase diagrams of these materials
may include multicritical points and also co-existence
of ferro- and antiferromagnetic phases, but the
role of Lifshitz-type couplings for mixed states is
difficult to establish for the electronic or structural
order parameters.
Hence, simple experimental systems displaying
mixed textures composed of different ordering modes have remained elusive.
The discovery of field-driven modulated magnetic state in
Ca$_3$Ru$_2$O$_7$ now provides an example of
a modulated state with mixed symmetries.

The phenomenological theory describing
its metamagnetic behavior is detailed in the section VI of the Supplemental materials.
The Landau-Ginzburg free energy displays
Lifshitz-type invariants
that are anisotropic in spin directions and
favour modulated coupled states between FM and AFM
spin-structure.
The specific form of these terms
reveals that they are caused by
spin-orbit interactions and encode the
twisting influence of the antisymmetric DM-exchange on
the magnetic order.

The observations demonstrate that metamagnetic crystals
with appropriate non-centrosymmetric structure
are an ideal playground for creating such textures.
As can be justified from the phenomenological Landau theory,
the spin-twisting DMIs in these magnets
preclude homogeneous phases, unless stabilized
by additional strong anisotropies, and
generically favour mixed AFM-FM states
near the metamagnetic transition.

FIG~\ref{PD4AFMs}c,d shows schematically possible magnetic
phase diagrams that can be realized in antiferromagnets
with non-centrosymmetric, in particular polar symmetry.

The tricritical region with first-order phase transitions
in the case of magnets with strong easy-axis anisotropy
is replaced by transitional regions covered by modulated mesophases.
Similarly, for systems with weak or absent anisotropies,
modulated states still can occur near the transition
towards the spin-polarized paramagnetic phase at elevated temperatures,
where antiferromagnetic and ferromagnetic order parameters
have similar magnitude and can become intertwined.
The transition from the paramagnetic state to the modulated
states is expected to be unconventional, implying inhomogeneous pre-cursor states.
The stabilization of this mixed magnetic texture relies on the unavoidable asymmetric exchange through spin-orbit couplings. It can be predicted for many systems to occur. So far, in Ca$_3$Ru$_2$O$_7$ we have observed only a mixed state
with a one-dimensional modulation. The basic mechanism that enables the generation
of this mixed state can act also in different spatial directions. Then, it may become
possible to create mixed textures similar to the chiral magnetic
skyrmions in non-centrosymmetric ferromagnets. Therefore, appropriate non-centrosymmetric metamagnets may also bear localized or
multidimensional lumps of one type order immersed in another one.  Phase transition involving such textures may allow to create condensates of
such lumps to form textured states that are simultaneously modulated in different directions,
akin to skyrmion lattices, but composed of different co-existing ordering modes.
We propose to call such states $\textit{improper Dzyaloshinskii textures}$, as they are composed of two ordering modes with different symmetry.

\section{Methods}

\subsection{Crystal growth and bulk characterization}
High quality single crystals of Ca$_3$Ru$_2$O$_7$ were grown using a floating zone method in a mirror furnace.
The single crystals were oriented using a white beam backscattering Laue X-ray diffraction method.
SFIG.5 in supplementary materials shows the corresponding Laue diffraction image indexed with
the Bb2$_1$m-structure and room temperature lattice parameters. The Laue diffraction
image shows sharp reflections, which indicate the excellent quality of the sample. The crystalline quality was further confirmed by measuring the rocking curve at (10,0,0) strong nuclear reflection in a neutron beam, SFIG.6. Measurements of the magnetization were performed using the vibrating sample magnetometer; specific heat was measured using the physical property measurements system by Quantum Design.
\subsection{X-ray diffraction and structure refinement}
As crystals of Ca$_3$Ru$_2$O$_7$ are easily cleaved, great care had to be taken to isolate a single crystal of adequate quality. Finally an irregular chip $(162\times  48\times  11\,\mu m^3)$ was selected and used in single crystal X-ray measurements on Rigaku AFC7 diffractometer with a Saturn 724+ CCD detector. After preliminary unit cell determination oscillation images around the unit cell axes proved good crystal quality without indications of partial cleavage or twinning, see SFIG.7. All diffraction experiments were performed at 295 K applying graphite-monochromated Mo-K$\alpha$ radiation ($\lambda$ = 0.71073 {\r{A}}) collimated with a mono-capillary. A total of three full $\varphi$-scans resulted in 2250 images from which after integration and scaling 7089 Bragg intensities were obtained. After averaging 1328 unique reflections were used in structure refinement. Derived lattice parameters, a = 5.3824(6) {\r{A}}, b = 5.5254(4) {\r{A}}, c = 19.5946(15) {\r{A}} (non-standard $\textit{Bb2$_1$m}$) are in a very good agreement with literature data. Refinement of the established model in space group $\textit{Cmc2$_{1}$}$ (standard setting of no. 36) converged in an excellent fit of 60 parameters vs. all 1328 independent reflections. Agreement based on F$^{2}$ including isotropic extinction correction is indicated by wR = 0.05 and goodness-of-fit = 1.12. However, a clear assignment of the absolute structure was not possible as refinement of a twin by inversion resulted in a volume ratio of 0.45(10) : 0.55. Based on careful analyses a centrosymmetric model as well as pseudo-tetragonal twinning had to be excluded. In an ongoing investigation we try to clarify if these observations point to a structural phase transition at high temperatures and are in accordance with anti-phase domain type features in optical micrographs. Details of diffraction experiment and structure determination are available from Cambridge Structural Database, CCDC 1901958.
\subsection{Small angle neutron scattering}
Two samples of a similar size from different growths were studied with SANS. We have carried out our SANS measurements on sample 1 using D33 SANS instrument at ILL in horizontal magnetic fields at temperatures between 10 K and 61 K. The measurements were performed using unpolarized neutrons with the wavelength $\lambda$=4.8 {\r{A}}. The neutron beam was collimated over 2.8 m before the sample. The sample to detector distance equaled 2 m. We have used the ILL $\emph{Blue Charly}$ 8 T horizontal magnet with the field oriented parallel to the neutron incident momentum. Typically, each scan was collected over 30 minutes to obtain a good statistics. The sample was cooled in a zero field to 2 K, at which the field was applied and the data was collected on heating. The temperature was then raised to 65 K (above $\textit{T}$$_{N}$ = 56 K), at which the field was reduced to zero, the sample was cooled to 2 K in zero field, then the next field was applied. SANS measurements on sample 2 were performed at the SANS-II instrument at PSI using a similar setup as at D33, with $\lambda$=4.93 {\r{A}}. Some of our earlier SANS measurements were performed at NG7 instrument at the NIST Center for Neutron Research.
\subsection{Transport and XMCD measurements}
For the transport measurements we prepared a microstructured device using a standard focused ion beam procedure. We fabricated a Hall bar device from an oriented single crystal of Ca$_3$Ru$_2$O$_7$ by the application of focused ion beam (FIB) as described elsewhere\cite{Ronning2017}. We cut a thin rectangular slice, with dimensions $100\times  20\times  3\,\mu m^3$ from the crystal and transferred it into non-conductive epoxy on a sapphire substrate. Ohmic contacts with approximately 10 $\Omega$ contact resistances were produced by sputter coating Au and annealing at $400^\circ$ C. The magnetoresistance measurements were performed in a LOT Quantum design magnet system.
XMCD measurements were performed on ID32 beamline at ESRF.
\subsection{Theoretical considerations}
Symmetry analysis and Landau-Ginzburg theories have been performed following standard procedures
\cite{ToledanoToledano1987} with input from the ISOTROPY program (http://stokes.byu.edu/iso/isowww.php).
Density-functional calculations have been performed using full-relativistic version of FPLO
(https://www.fplo.de/) \cite{Koepernik1999}.

\section{Acknowledgements}
We thank U. Nitzsche for technical support with FPLO. N. K. acknowledges the support from JSPS KAKNHI (No. JP17H06136). D. A. S. thanks C. Geibel for the critical reading of the manuscript and constructive comments. Access to NG7 SANS was provided by the Center for High Resolution Neutron Scattering, a partnership between the National Institute of Standards and Technology and the National Science Foundation under Agreement No. DMR-1508249. We thank J. Krzywon and Y. Qiang for technical support during SANS experiment at NIST. This work is partly based on experiments performed at the Swiss spallation neutron source SINQ, Paul Scherrer Institute, Villigen, Switzerland.

\section{Author contributions}
U.K.R. conceived the project. U.K.R., A.P.M. and D.A.S. supervised the project. N.K. and D.A.S. grew single crystals. D.A.S oriented and characterised samples. H.B. and U.B. analysed the crystal structure. T.H. performed the electrical transport measurements and analysed the Hall effect data. K.K. performed XMCD measurements. D.A.S., R.C., J.S.W., and M.B. performed SANS measurements. D.A.S. and E.R. carried neutron diffraction experiments. U.K.R. carried out DFT calculations and developed the Landau-Ginzburg-type free energy theory. D.A.S. and U.K.R wrote the manuscript with contributions from all co-authors.

\section{Data availability} The data that support the plots within this paper may be downloaded at\cite{Sokolov2018a}. The datasets for the small-angle neutron scattering experiments
on D33 are available from the Institute Laue-Langevin data portal (10.5291/ILL-DATA.5-42-462)\cite{Sokolov2018}.

\end{document}


\title{
%
Supplementary Information for Metamagnetic texture in a polar antiferromagnet
%
%
}

\author{D. A. Sokolov}
\affiliation{Max-Planck-Institut f\"{u}r Chemische Physik fester Stoffe, D-01187 Dresden, Germany}

\author{N. Kikugawa}
\affiliation{National Institute for Materials Science, Tsukuba 305-0047, Japan}

\author{T. Helm}
\affiliation{Max-Planck-Institut f\"{u}r Chemische Physik fester Stoffe, D-01187 Dresden, Germany}

\author{H. Borrmann}
\affiliation{Max-Planck-Institut f\"{u}r Chemische Physik fester Stoffe, D-01187 Dresden, Germany}

\author{U. Burkhardt}
\affiliation{Max-Planck-Institut f\"{u}r Chemische Physik fester Stoffe, D-01187 Dresden, Germany}

\author{R. Cubitt}
\affiliation{Institut Laue-Langevin, 6 Rue Jules Horowitz, F-38042 Grenoble, France}

\author{J. S. White}
\affiliation{Laboratory for Neutron Scattering and Imaging (LNS), Paul Scherrer Institute (PSI), CH-5232 Villigen, Switzerland}

\author{E. Ressouche}
\affiliation{Univ. Grenoble Alpes, CEA, INAC-MEM, 38000 Grenoble, France}

\author{M. Bleuel}
\affiliation{NIST Center for Neutron Research, NIST, Gaithersburg, MD 20899, USA}

\author{K. Kummer}
\affiliation{ESRF, 71 avenue des Martyrs, 38000 Grenoble, France}

\author{A. P. Mackenzie}
\affiliation{Max-Planck-Institut f\"{u}r Chemische Physik fester Stoffe, D-01187 Dresden, Germany}
\affiliation{Scottish Universities Physics Alliance (SUPA), School of Physics and Astronomy, University of St Andrews, St Andrews KY16 9SS, United Kingdom}

\author{U.K. R\"o\ss ler}
\affiliation{IFW Dresden, PO Box 270116, D-01171 Dresden, Germany}

\maketitle

\section{X-ray magnetic circular dichroism (XMCD)}
The X-ray magnetic circular dichroism (XMCD) experiment was performed at the ESRF beamline ID32 using the high-field magnet endstation\cite{kummer16}. The samples were cleaved in situ in the high field magnet at low temperature and a pressure better than $2\times10^{-10}\,$mbar leaving behind a clean $ab$ surface. The magnetic field was applied along the beam direction. The $b$ axis of the samples was titled by $15^\circ$ towards the $c$ axis with respect to the field direction, resulting in a small field component along $c$ but zero field along $a$.  The X-ray absorption spectra were taken in total electron yield mode with circular left and circular right polarization for both positive as well as negative field direction. The XMCD was determined from all four spectra as the difference between spectra taken with opposite helicity.

\begin{figure}
\centering
\includegraphics[width=1.0\columnwidth]{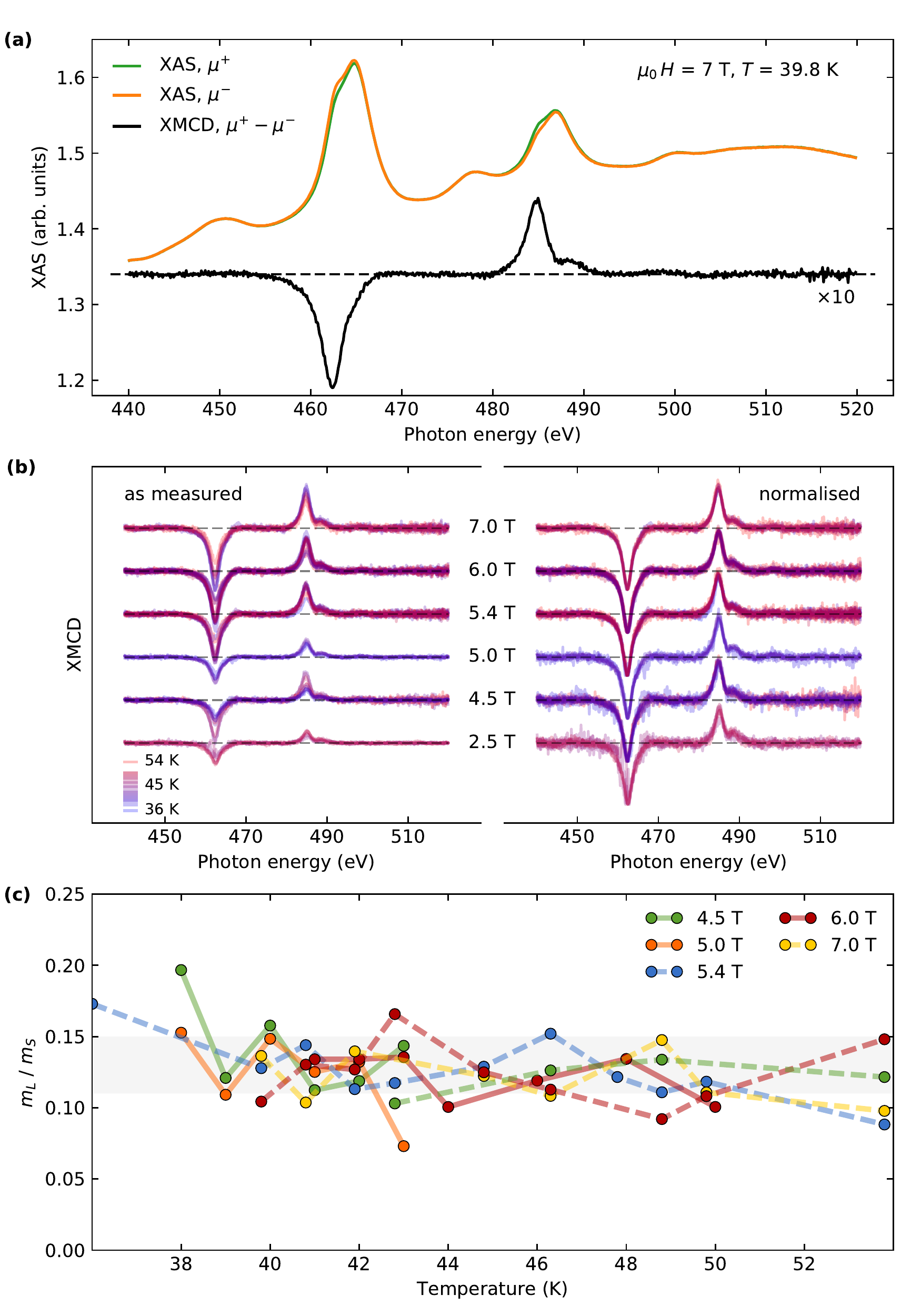}
\caption{(a) X-ray absorption spectra for both experimental helicities and the corresponding XMCD in the CAFM phase. (b) XMCD as a function of field and temperature in the relevant region of the phase diagram. The XMCD intensities are shown both as measured to ease comparison with macroscopic magnetization measurements (left) and normalised to average intensity to better compare the lineshapes between the XMCD spectra (right). (c) Ratio of orbital to spin moment as extracted from the XMCD data.}
\label{xmcd}
\end{figure}

Fig.~\ref{xmcd}a shows the XAS spectra for positive and negative helicity and the XMCD obtained as the difference of the two in the CAFM phase at low temperature and high field. In addition to the Ru $M_3$ and $M_2$ absorption lines there is an additional broad X-ray absorption features around 450\,eV which displays no XMCD and is due to Ca $L_{3,2}$ absorption. We followed the XMCD signal at the Ru $M_{3,2}$ as a function of magnetic field and temperature in the relevant region of phase space. The obtained spectra are shown in Fig.~\ref{xmcd}b both with the relative intensities as measured (left panel) and with the intensities normalized to peak value (right panel). The intensity of the XMCD signal (left panel) scales well with the macroscopic magnetization curves shown in Fig. 2a of the manuscript. The lineshape of the XMCD, however, does not change, neither with field nor temperature, across the meta-magnetic, magnetic and MIT transitions. Using the XMCD sum rules, we can extract the ratio of orbital to spin moment aligned along the field which only depends on the ratio of the integrated XMCD signal at the $M_3$ and $M_2$ absorption edge, respectively\cite{thole92}. As the XMCD lineshape does not change across the phase diagram we always find the same value $<m_L>/<m_S>\approx 0.13\pm0.02$, confirming the presence of a sizable orbital moment. In principle, spin- and orbital moments can also be extracted individually from the XMCD signals\cite{thole92}. We refrained from doing so here because of the complex background of the XAS spectra which makes it difficult to extract reliable numbers.

The robustness of the XMCD lineshape and the $<m_L>/<m_S>$ ratio suggests that the local crystal field experienced by the Ru ions in the RuO$_6$ octahedra does not change significantly in the interesting region of the phase diagram.

\section{Transport measurements}
\begin{figure*}[ht]
\includegraphics[scale=0.45]{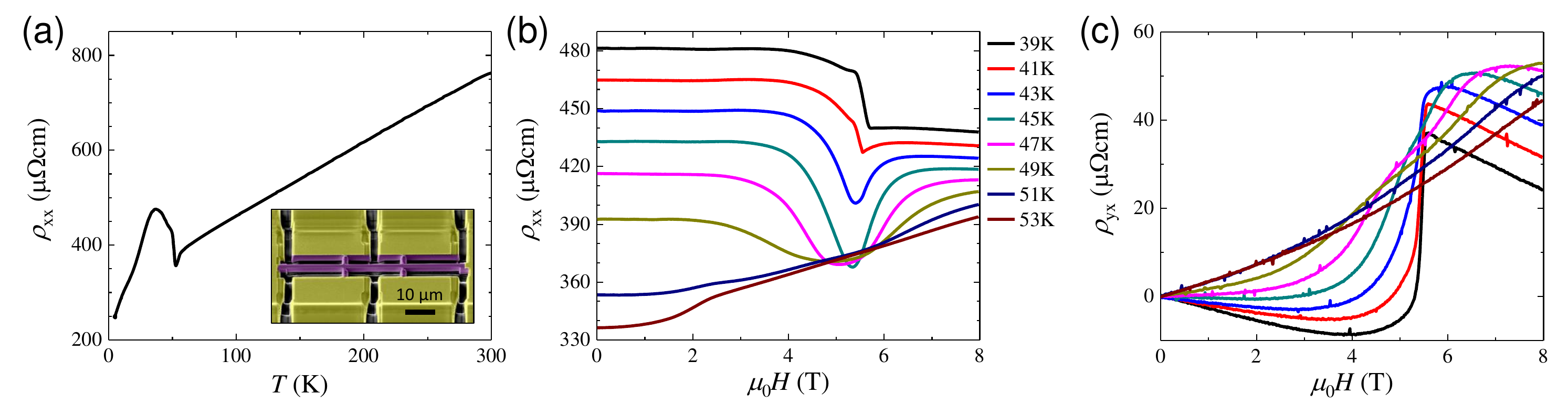}
\caption{\label{fig:Metamagnetic}Transport measurements on a microstructured single crystal of Ca$_3$Ru$_2$O$_7$: (a) Temperature dependence of the in-plane resistivity. Inset: False color SEM image of the FIB-microfabricated Hall-bar device (purple) with gold contacts (yellow). Current runs along the a-axis. (b) In-plane magnetoresistivity and (c) Hall resistivity curves at various temperatures for field applied along the b-axis.}
\label{HALL1}
\end{figure*}

\begin{figure*}[ht]
\includegraphics[scale=0.45]{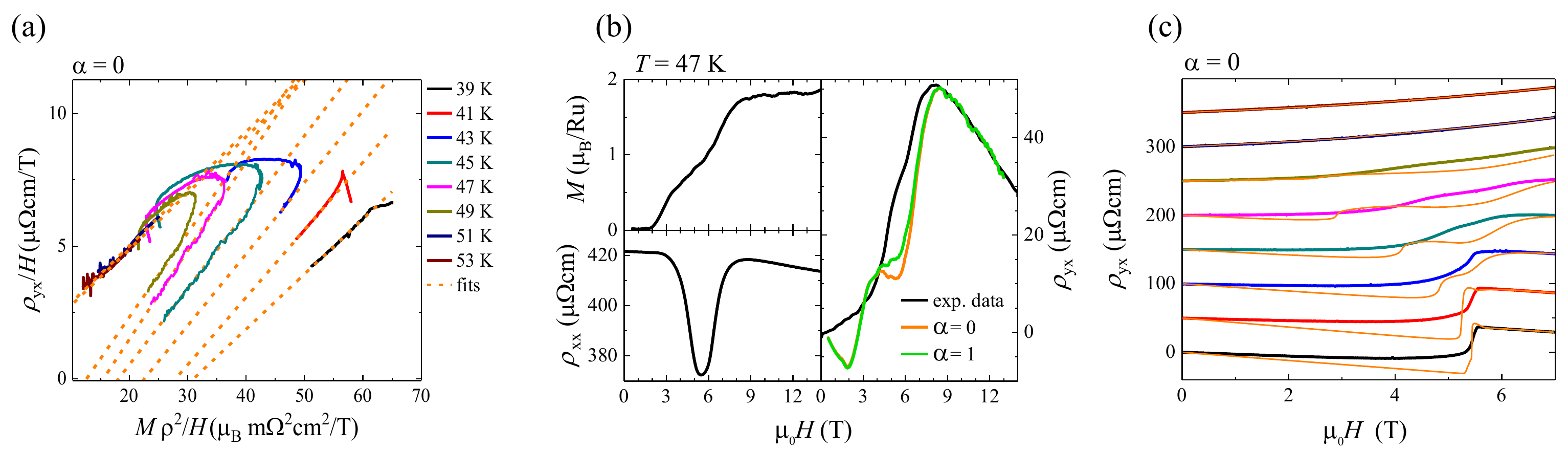}
\caption{\label{fig:Metamagnetic}(a) Linear fits (orange lines) to the Hall data from Fig~\ref{HALL1} (c) to Eq. 2 as described in the text. The respective fit parameters we show in Table 1. (b) Example curves (black) of magnetization, in-plane resistivity and Hall resistivity, at 47 K used for fitting the Hall data according to Eq. 2. Pink and red curves are two simulations according to Eq. 2 with $\alpha$=0,1, respectively. (c) Hall resistivity data from Fig~\ref{HALL1}c, offset for better visibility. Orange curves are simulations according to Eq. 2 with $\alpha$=0 using parameters from Table 1, see text for further details.}
\label{HALL2}
\end{figure*}

We carried out magnetotransport measurements on a sample in the Hall-bar geometry, prepared by focused Ion beam (FIB) microfabrication. A lamella with dimensions (3$\times$20$\times$100) $\mu$m$^{3}$ was cut from a single-crystal using Gallium FIB. The device is shown in the inset of Fig~\ref{HALL1}a. We used a standard four-terminal Lock-In method for measurements of electrical resistivity. We applied a current of 100 $\mu$A with frequency f = 177 Hz along the $a$-axis while the magnetic field was applied along the $b$ axis. The zero-field resistivity curve shown in Fig~\ref{HALL1}a resembles data reported previously\cite{kikugawa2010} and demonstrates the high quality of micromachined devices. The in-plane resistivity, $\rho_{xx}$, as well as the Hall resistivity, $\rho_{yx}$, exhibit a step-like behavior at low temperatures (see Fig~\ref{HALL1}b and c). The sharp change evolves into a broader transition as temperature approaches 50 K. Above 50 K, both $\rho_{xx}$ and $\rho_{yx}$ follow an overall positive slope. Most interestingly, we observe a two-step-like behavior in $\rho_{yx}$ for temperatures between 45 K and 49 K, the range in which the metamagnetic texture was observed in small-angle neutron scattering measurements. In general, the Hall resistivity can be a composition of three components:

\begin{equation}
\rho_{yx} = \rho_{yx}^{N} + \rho_{yx}^{A} + \rho_{yx}^{T},
\end{equation}
where N, A, T denote the normal, anomalous and topological Hall effect contributions\cite{nagaosa2010}. Using the expression:

\begin{equation}
\frac{\rho_{yx}}{H} = R_{0} + \frac{S_{A}\rho_{xx}^{\alpha}M}{H} + \frac{\rho_{yx}^{T}}{H},
\end{equation}

where H is the magnetic field and M is the magnetization, we can extract the normal and anomalous Hall coefficients,  R$_{0}$ and S$_A$ (see Fig~\ref{HALL2}a). The respective intercept and slope of the linear fits to the high-field part of $\rho_{yx}$/H plotted versus M$\rho_{xx}^{2}$/H are listed in Table 1.
Again using these materials parameters and  $M(H)$ data, the Hall resistivity  $\rho_{yx}$ can be \textit{simulated}
under assumption that only the normal and anomalous contributions exist.
In Fig~\ref{HALL2}b we show the simulation result for T = 47 K compared to the raw data, M, $\rho_{xx}$ and $\rho_{yx}$.
Both case $\alpha$=0 and 1 are shown, which correspond to an anomalous contribution
dominated by intrinsic or extrinsic scattering mechanisms, respectively \cite{shiomi2012}.
As can be seen from Fig~\ref{HALL2}c deviations between simulated Hall resistivity and
the experimental data appear for temperatures below 49 K. This indicates the existence of a so-called topological
contribution to $\rho_{yx}$ in the state corresponding to the $H-T$-region of the phase diagram, where the metamagnetic texture
is observed. It may be related to either a topological contribution or another non-trivial contribution
that is not considered through normal and anomalous Hall resistivity. In particular, it may derive from effects of
a non-collinear spin-structure \cite{shiomi2012}.

\begin{table}[h!]
\centering
\begin{tabular}{||c c c||}
 \hline
 T & R$_{0}$ & S$_{A}$ \\ [0.5ex]
 \hline\hline
 39 & -6.4 & 2.1e-4 \\
 \hline
 41 & -7.1 & 2.5e-4 \\
 \hline
 43 & -6.1 & 2.7e-4 \\
 \hline
 45 & -5.1 & 2.8e-4 \\
 \hline
 47 & -5.2 & 3.3e-4 \\
 \hline
 49 & -4.0 & 3.2e-4 \\
 \hline
 51 & 0.7 & 2.2e-4 \\
 \hline
 53 & 0.63 & 2.1e-4 \\ [1ex]
 \hline
\end{tabular}
\caption{Linear fit parameters extracted from Fig~\ref{HALL2}a.}
\label{table:1}
\end{table}

\section{Small-angle neutron scattering measurements}

\begin{figure*}[ht]
\includegraphics[scale=0.5]{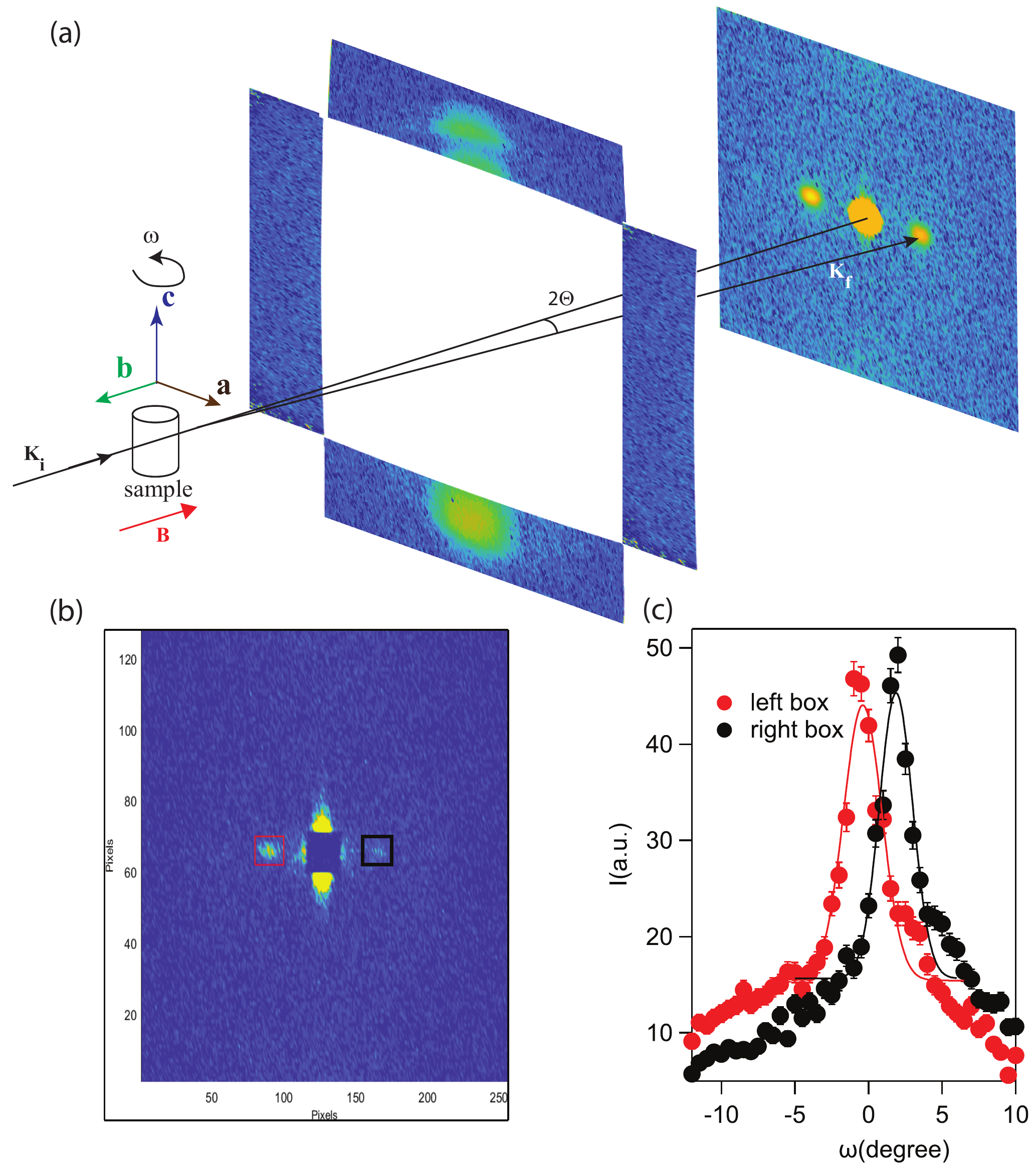}
\caption{\label{fig:Metamagnetic}(a) Geometry of SANS experiment on D33 instrument at ILL. (b) Raw SANS pattern collected at 48 K in magnetic field of 2 T. (c) Rocking scan of magnetic Bragg peaks shown in (b). Intensities in the left and right boxes in panel (b) are indicated by red and black markers, respectively. Solid lines are best fits to Gaussian lineshapes.}
\label{suppl_sans}
\end{figure*}

Our SANS sample 1 measured at ILL was a 238 mg single crystal, mounted on Al sample holder with the c-axis vertical and the b-axis parallel to the field and the neutron incident momentum. The approximate sample dimensions along the major crystallographic axes were a=8.3 mm, b=7.7 mm, c=1.5 mm. The SANS sample 2 measured at PSI was a 215 mg single crystal with approximate dimensions of a=6 mm, b=5.3 mm, c=2.8 mm. The sample 2 was measured in the same orientation as sample 1. $\xi_{b}$, the correlation length of the magnetic texture along the $\textit{b}$-axis, directed along the incident momentum of neutrons was calculated from the rocking curve measurement (rotation around the c-axis), schematically illustrated as $\omega$ rotation in Fig~\ref{suppl_sans}a. The rocking curve at 2 T and 48 K, obtained by rotating the sample together with a cryomagnet was fitted by a Gaussian lineshape, which yielded FWHM=3.1 and 2.7 degrees for the satellites contained in left and right boxes, Fig~\ref{suppl_sans}b,c.

\begin{figure*}[ht]
\includegraphics[scale=0.5]{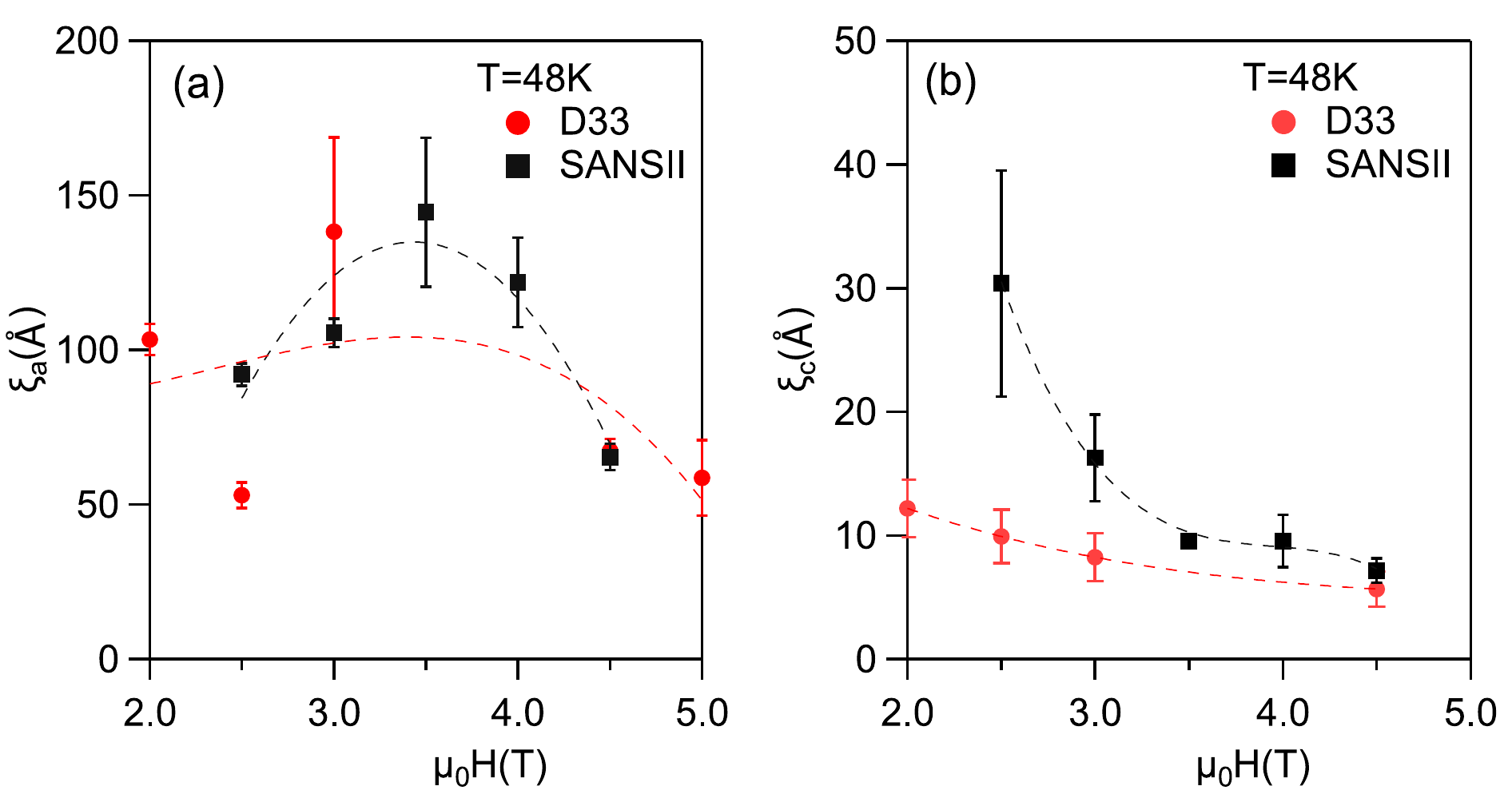}
\caption{\label{fig:Metamagnetic}Magnetic field dependence of the correlation length along the $\textit{a}$ (a) and $\textit{c}$-axis (b) at 48 K. Dashed lines are a guide to the eye.}
\label{sans2}
\end{figure*}

\begin{equation}
\xi_{b}=\frac{8\textrm{ln}2}{Q \textrm{tan}(\textit{b})},
\end{equation}

where $\textit{b}$ is the full width at half maximum of the rocking curve and Q is the wavevector of the observed reflection.  For estimates of $\xi_{a}$ and $\xi_{c}$, the correlation lengths along the $\textit{a}$ and $\textit{c}$ axes we used the tangential and radial widths of the satellites, which were fitted to Gaussian lineshapes. The experimental resolution in a typical SANS geometry is largely determined by the wavelength spread, $\Delta$$\lambda/\lambda$ and d-spacing spread of a sample, $\Delta$d/d. For the estimates of $\xi_{b}$ we ignored the effects of the instrumental resolution since $\textit{b}>>\Theta\Delta\lambda/\lambda$ and the width of the rocking curve is much greater than the angular size of the direct beam ($\sim$0.44 degree). Here 2$\Theta$ is the scattering angle.
\begin{equation}
\xi_{a}=\frac{1}{Q(\Delta d/d)},
\end{equation}
where $\Delta$d/d is the d-spacing spread of the lattice, which is estimated from the radial width, R$_{w}$ using the following,
\begin{equation}
(R_{w}/2\Theta)^{2}=(\textit{a}/2\Theta)^{2}+(\Delta d/d)^{2}+(\Delta\lambda/\lambda)^{2},
\end{equation}
where $\textit{a}$ is the angular size of the direct beam as detailed in Ref.\cite{cubitt92}.
\begin{equation}
\xi_{c}=\frac{8\textrm{ln}2}{Q \textrm{tan}(\textit{t/2})},
\end{equation}
where $\textit{t}$ is the intrinsic tangential variation of the lattice, which is estimated from the tangential widths of the reflection, $\textit{T$_{w}$}$ and direct beam, $\textit{az}$ using the following,
\begin{equation}
T_{w}=\sqrt{t^{2}+\textit{az}^{2}},
\end{equation}
The field dependence of $\xi_{a}$ and $\xi_{c}$ is shown in Fig~\ref{sans2}. Whereas no clear field dependence was observed for $\xi_{a}$, a moderate suppression of $\xi_{c}$ by the field was evidenced on both instruments. In our SANS geometry the resolution in the detector plane (our $\textit{a}$-, and $\textit{c}$-axes) is approximately one of order of magnitude lower than along the neutron flux direction (our $\textit{b}$-axis). Therefore, the values of $\xi_{a}$ and $\xi_{c}$ should be regarded as a lower limit on correlation length in the $\textit{ac}$ plane.

\section{Single crystal growth and orientation of samples}
Single crystals of Ca$_3$Ru$_2$O$_7$ were grown using a floating zone method in a mirror furnace (Canon Machinery, model SCI-MDH)), as reported elsewhere\cite{perry04}. The crystal growth was performed in an atmosphere of the mixture of Ar and O$_{2}$ (Ar:O$_{2}$=85:15).

\begin{figure*}[ht]
\includegraphics[scale=0.350]{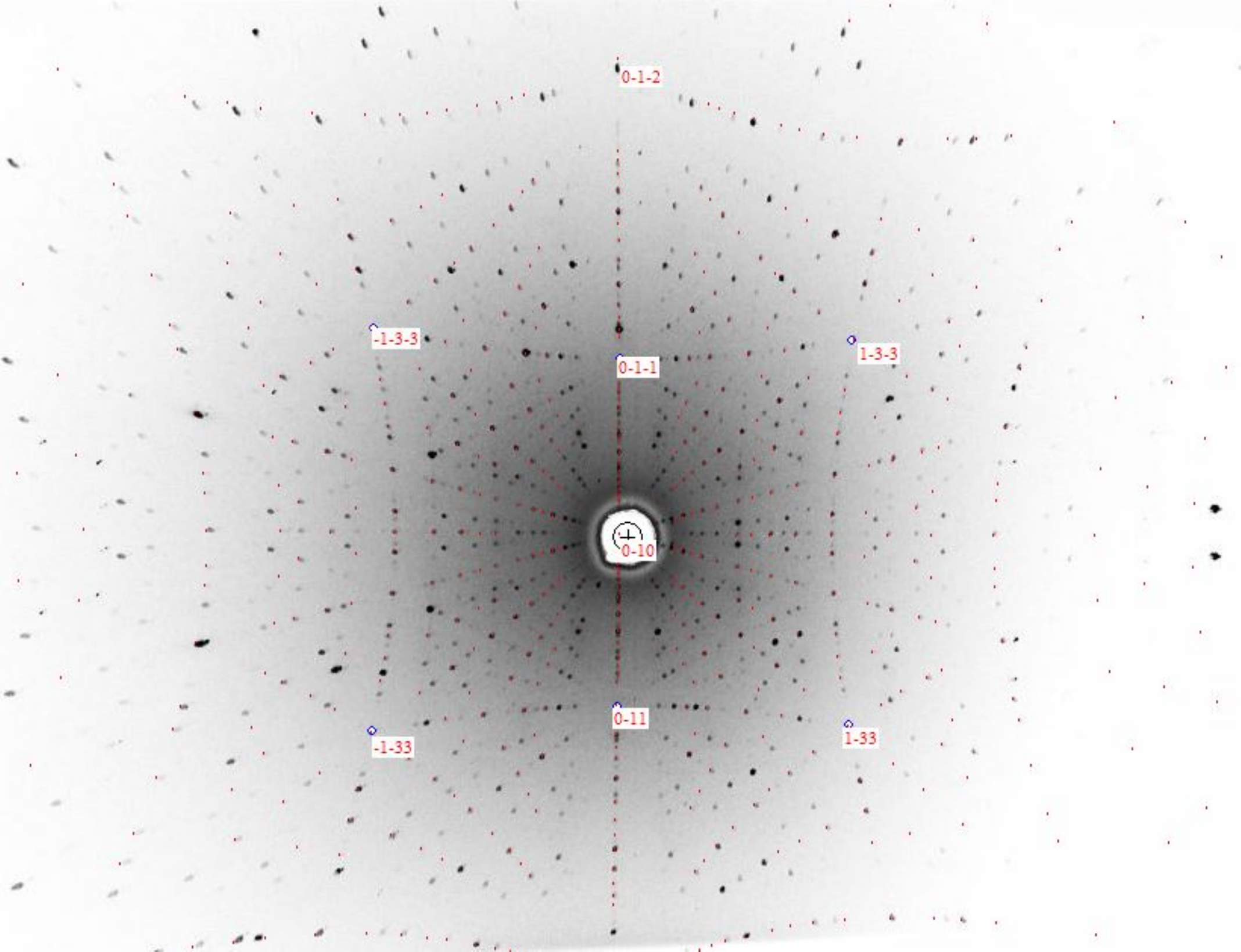}
\caption{\label{fig:Metamagnetic}The room temperature X-ray Laue diffraction pattern of an as-grown (010) facet of a Ca$_3$Ru$_2$O$_7$ single crystal. The red spots and assigned Miller indices show the calculated diffraction pattern of the Bb2$_{1}$m orthorhombic-space group.}
\label{010}
\end{figure*}
\begin{figure*}[ht]
\includegraphics[scale=0.6]{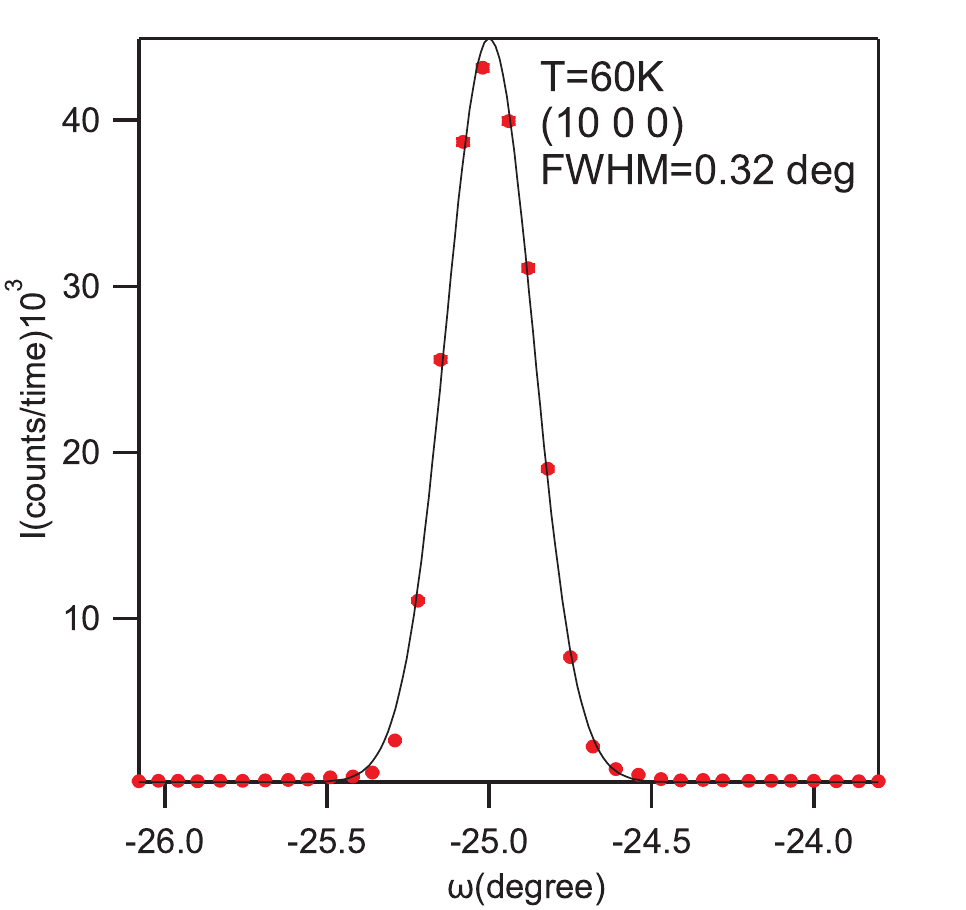}
\caption{\label{fig:Metamagnetic}The rocking curve at (10 0 0) reflection measured in a thermal neutron beam. The peak is fitted by a Gaussian lineshape with the full width at the half maximum FWHM=0.32 degree.}
\label{ROCKING}
\end{figure*}

The single crystals were oriented using the X-ray Laue backscattering method utilising a home-built instrument. The typical pattern shown in Fig~\ref{010} demonstrates very sharp reflections and allows to distinguish between the $\textit{a}$- and $\textit{b}$-axes. The full width at the half maximum (FWHM)=0.32 degree of the rocking curve measured at the strong nuclear reflection (10 0 0) in a neutron beam with the wavelength $\lambda$=1.272 {\r{A}} on D23 instrument at ILL indicated an excellent quality of our crystal, Fig~\ref{ROCKING}.

\begin{figure*}[ht]
\includegraphics[scale=0.6]{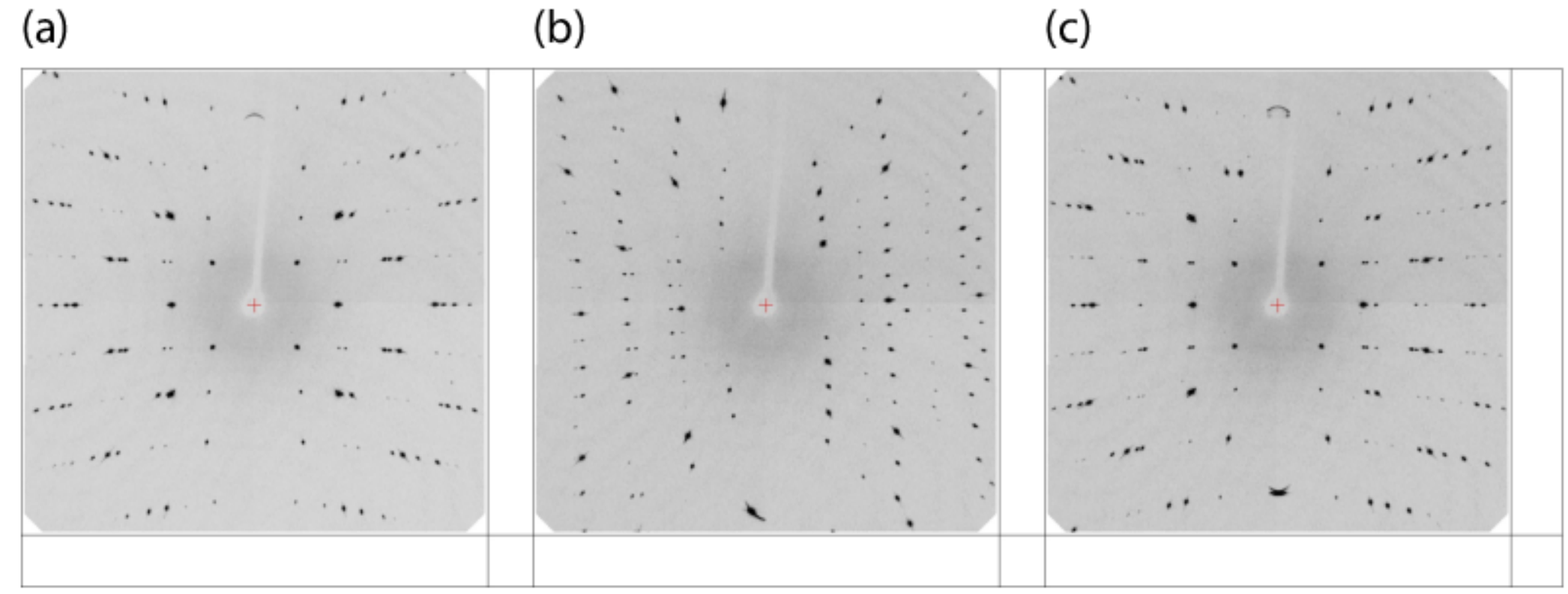}
\caption{\label{fig:Metamagnetic}(a), (b) and (c) show X-ray diffraction patterns of
a small Ca$_3$Ru$_2$O$_7$ crystal after rotation about the crystallographic $\textit{a}$, $\textit{b}$, and $\textit{c}$-axis respectively. The
rotation axis in each case is vertical.}
\label{XRAY}
\end{figure*}

Oscillation images around crystallographic axes confirm that crystal quality was maintained for the small single crystal, Fig~\ref{XRAY}.

\section{DFT calculations}
We have ascertained the
presence of a sizeable effect of spin-orbit couplings (SOC) and orbital magnetic
moments by standard DFT-calculations within the full-potential local orbital (FPLO)
approach \cite{Koepernik1999}. We used the generalized gradient approximation
(GGA) as exchange-correlation functional\cite{PBE1996}.
Correlations beyond have been included by the GGA+U method
for the 4d-states of Ru with an effective Hubbard-like U=0 to 3~eV.
The fully relativistic FPLO code includes SOC to all orders,
being based on solutions of the 4-spinor Kohn-Sham-Dirac equations.
As a relevant example, for ferromagnetic spin configurations
we find spin-moments $m_s=$~1.56~$\mu_B /$ Ru ion and
orbital moments $m_o=$~0.19~$\mu_B /$ Ru ion for a representative
value of U = 2.25~eV. However, there are large
uncertainties regarding exact values of spin and orbital moments,
as seen from calculated results in Table~\ref{table:2}
and the appropriate values for the DFT+U-correction
in the metallic state at temperatures above the metal-insulator transition
are uncertain. For $U \ge 3$~eV a gap opens and the
bandstructure would correspond to an insulating ground-state,
in good agreement with earlier DFT-results by Liu \cite{Liu2011}.
However, the results indicate the presence of relatively strong SOC effects
in the collinear fully polarized state. This also suggests
that antisymmetric Dzyaloshinskii-Moriya exchange is relatively strong
in Ca$_3$Ru$_2$O$_7$.
In view of the complex bi-layer structure of Ca$_3$Ru$_2$O$_7$
and its correlated metallic character at the relevant
higher temperatures, a credible microscopic evaluation of the
Dzyaloshinskii-Moriya interactions (DMIs) is hardly feasible.
However, as the Ru ions occupy the general 8b Wyckoff positions
in Ca$_3$Ru$_2$O$_7$, the microscopic DMIs between the
spins $\mathbf{s}_i$ on these sites,
$\mathbf{D}_{ij}\cdot(\mathbf{s}_i \times \mathbf{s}_j)$
are allowed for all pairs of sites with a general
Dzyaloshinksii vector $\mathbf{D}_{ij}$, which is only
determined by the SOC in the spin-split electronic bandstructure.

\begin{table}[h!]
\centering
\begin{tabular}{||c c c||}
 \hline
 U  & $m_{s}$ & $m_{o}$ \\ [0.5ex]
 eV & $\mu_B/$~Ru  &  \\ [0.5ex]
 \hline
 \hline
 0    & 1.53 & 0.09 \\
 1.00 & 1.46 & 0.12 \\
 1.50 & 1.53 & 0.09 \\
 2.00 & 1.52 & 0.20 \\
 2.25 & 1.56 & 0.19 \\
 3.00 & 1.74 & 0.02 \\ [1ex]
 \hline
\end{tabular}
\caption{Magnetic spin moment $m_s$ and orbital moment $m_o$ on Ru in
 Ca$_3$Ru$_2$O$_7$ from GGA+U density functional theory calculations.}
\label{table:2}
\end{table}

\section{Landau-Ginzburg free energy}
The primary magnetic order in Ca$_3$Ru$_2$O$_7$
has been identified
as a simple antiferromagnetic two-sublattice structure,
where ferromagnetically coupled Ru-bilayers alternate
with antiparallel moments stacked in $c$-direction \cite{yoshida05}.
The antiferromagnetic order breaks the B-centering
operation with the vector $t =  (1/2, 1/2, 0)$ (in
the Cartesian coordinate system of Fig.~4~a, which will
be used for spatial and spin-coordinates in the following).
The metamagnetic behavior of this
two-sublattice order, and the tricritical behavior Fig.~4~c,
can be described by the Landau theory
for the coupling of the two equivalent sublattices
$\mathbf{M}_{\textrm{I}}$ and  $\mathbf{M}_{\textrm{II}}$,
but this expansion requires higher-order terms \cite{Tuszynski1981}.
In the vicinity of the tricritical point
the expansion can be expressed by
a free energy density
\begin{eqnarray}
F_0 & = & B_2\,( |\mathbf{M}_{\textrm{I}}|^2
+ |\mathbf{M}_{\textrm{II}}|^2 )  \nonumber \\
& + & A_2 \mathbf{M}_{\textrm{I}}\cdot\mathbf{M}_{\textrm{II}}  \nonumber \\
& + & B_4 ( |\mathbf{M}_{\textrm{I}}|^4
+ |\mathbf{M}_{\textrm{II}}|^4 )  \nonumber \\
& + & A_4 (\mathbf{M}_{\textrm{I}}\cdot\mathbf{M}_{\textrm{II}})^2
\nonumber \\
& + & B_6 ( |\mathbf{M}_{\textrm{I}}|^6
+ |\mathbf{M}_{\textrm{II}}|^6 )  \,.\\
\end{eqnarray}
Representing this phenomenological theory in terms of
the staggered vector
$\mathbf{l}=(1/2)(\mathbf{M}_{\textrm{I}}-\mathbf{M}_{\textrm{II}})$
of antiferromagnetism and the net magnetic moment
$\mathbf{f}=(1/2)(\mathbf{M}_{\textrm{I}}+\mathbf{M}_{\textrm{II}})$
leads to a free energy, which should include
at least 6$^{th}$ order terms in the Landau expansion to
describe the tricritical point and the phase coexistence
between antiferromagnetism and ferromagnetic field-enforced states,
\begin{eqnarray}
 w_0  & = & a_l | \mathbf{l} |^2 + a_f | \mathbf{f} |^2 \nonumber \\
      & + & b_l | \mathbf{l} |^4 + b_f | \mathbf{f} |^4 \nonumber \\
      & + & c_1 | \mathbf{l} |^2 \, | \mathbf{f} |^2  \nonumber \\
      & + & c_l  | \mathbf{l} |^6 + c_f | \mathbf{f} |^6 \nonumber \\
      & + & c_2  | \mathbf{l} |^4 | \mathbf{f} |^2
          + c_3  | \mathbf{l} |^2 | \mathbf{f} |^4  \,.
\end{eqnarray}
The two co-existing symmetry modes $\mathbf{l}$ and $\mathbf{f}$
and their Cartesian spin-components belong to odd and even
representations of the Cmc2$_1$ space-group (which is a standard setting of space group No36 equivalent to Bb2$_1$m) with respect to the
partial $t$-translation, i.e. they have different symmetry.
As this crystal lattice of Ca$_3$Ru$_2$O$_7$
has a non-centrosymmetric orthorhombic symmetry belonging
to Laue class 2mm, Landau-Ginzburg free energies
for these two modes can have Lifshitz invariants,
i.e. terms linear in spatial gradients of Cartesian
components of either of these modes.
These terms derive from the Dzyaloshinskii-Moriya
interactions and can be written as
combinations of bilinear antisymmetric forms,
\begin{equation}
\Gamma_{ij}^{(\gamma)} (\mathbf{x}) \equiv
( x_i\, \partial_{\gamma}x_j - x_j \, \partial_{\gamma}x_i) \,.
\end{equation}

For the 2mm symmetry and the simple antiferromagnetism
in  Ca$_3$Ru$_2$O$_7$, the corresponding free energy
contributions are
\begin{eqnarray}
w_D & = & D_x\,\Gamma_{zx}^{(x)}(\mathbf{l})
        + D_y\,\Gamma_{yy}^{(y)}(\mathbf{l}) \\
w_F & = & F_x\,\Gamma_{zx}^{(x)}(\mathbf{f})
        + F_y\,\Gamma_{yz}^{(y)}(\mathbf{f}) \,,
\end{eqnarray}
where the coefficients $D_{x,y},F_{x,y}$ are materials constants.
These contributions, in particular $w_D$ lead to
the spiralling cycloidal
modulations of the magnetic order \cite{Dzyaloshinskii1964},
which we call \textit{Dzyaloshinskii textures}.
In an antiferromagnet where
only the $w_D$ invariants are acting, an antiferromagnetic
spiral would be composed only of the $\mathbf{l}$-symmetry mode.
Therefore, we can refer to such a sprial as a ``proper'' texture.
In the schematic phase diagrams, Fig.4c and Fig.4d, the
presence of term $w_D$ can lead to spiralling or
other precursor textures designated $\pi_1$ in particular
at higher temperatures above the temperature range,
where the  anisotropies enforce a homogeneous antiferromagnetic state,
which seems to be the case in Ca$_3$Ru$_2$O$_7$.
Otherwise, the $w_D$ term can affect
the antiferromagnetic order-parameter and
could lead to a spiralling antiferromagnetic ground-state for weak enough anisotropies.
The existence of these terms also means that the thermal phase transition from the
paramagnetic to the antiferromagnetic state does not
obey the Lifshitz criterion of the Landau theory for
a continuous phase transition. Therefore, the thermal
ordering transition in a material with a contribution of
the form $w_D$ is expected to be anomalous. In particular,
a fluctuating precursor states can arise above the magnetic ordering.

A complete phenomenological theory then requires also the
usual squared gradient terms of the order-parameter,
\begin{eqnarray}
w_E & = &  A_l \, (\nabla \mathbf{l})^2  +
           A_f \, (\nabla \mathbf{f})^2  + \dots \,,
\end{eqnarray}
where the ellipses stand for anisotropic exchanges terms.
Indeed, in  Ca$_3$Ru$_2$O$_7$ strong additional anisotropies,
suppress a modulation in the antiferromagnetic ground state.
A complete Landau theory for the homogeneous states would require
additional terms, the leading magnetocrystalline anisotropy
being described by
\begin{eqnarray}
w_a & = & K_z\,l_z^2 + k_z\,f_z^2 \nonumber  \\
    & + & \kappa_x\, l_x^2 + \kappa_{xy}\, l_x\,l_y + \kappa_y \, l_y \nonumber \\
    & + & \nu_x\, f_x^2 + \nu_{xy}\, l_x\,l_y + \nu_y \, l_y \,,
\end{eqnarray}
with anisotropy coefficients $K_z,\kappa,\nu$.
We note that for doped compounds Ca$_3$(Ru$_{1-x}$TM$_x$)$_2$O$_7$,
where TM is Fe or Mn, incommensurately modulated antiferromagnetic
ground-states have been observed  \cite{Ke2014,Zhu2017}
and have been interpreted as Dzyaloshinskii spirals \cite{Zhu2018}.
The observation suggests that the substitituion on the magnetic
site weakens the anisotropy and reveals the presence of
the inhomogeneous terms $w_D$, such that the $w_D$ terms overcomes
the anisotropies $w_a$.

In the region of the metamagnetic phase co-existence, additional
higher order Lifshitz terms become operative, which couple
$\mathbf{l}$ and $\mathbf{f}$.
There are a great number improper of couplings between these
modes in  Ca$_3$Ru$_2$O$_7$.
With the aim to illustrate the complexities of possible
effects, we give here a complete list of these terms.
The mixed higher order terms are Lifshitz-type invariants
as follows:
\begin{equation}
w_{\mu}  =
\sum_{\alpha=x,y,z}\,\sum_{\beta=x,y}
\left(a_{\alpha}\,f_{\alpha}^2\,\Gamma_{\beta \, z}^{(\beta)})(\mathbf{l})
    + b_{\alpha}\,l_{\alpha}^2\,\Gamma_{\beta \, z}^{(\beta)}(\mathbf{m})
\right)
\end{equation}
and

\begin{eqnarray}
w_{\Delta} & = &
       \Delta_1 f_x f_y \Gamma_{xy}^{(z)}(\mathbf{l}) \nonumber \\
 & + & \Delta_2 f_x f_y \Gamma_{zx}^{(z)}(\mathbf{l}) \nonumber \\
 & + & \Delta_3 f_x f_y \Gamma_{yz}^{(x)}(\mathbf{l}) \nonumber \\
 & + & \Delta_4 f_x f_y \Gamma_{yz}^{(z)}(\mathbf{l}) \nonumber \\
 & + & \Delta_5 f_x f_y \Gamma_{zx}^{(y)}(\mathbf{l}) \nonumber \\
 & + & \Delta_6 f_z f_x \Gamma_{xy}^{(y)}(\mathbf{l}) \nonumber \\
 & + & \Delta_7 f_z f_x \Gamma_{zx}^{(z)}(\mathbf{l}) \nonumber \\
 & + & \Delta_8 f_y f_z \Gamma_{xy}^{(x)}(\mathbf{l}) \nonumber \\
 & + & \Delta_9 f_y f_z \Gamma_{yz}^{(z)}(\mathbf{l}) \nonumber \\
 & + & \Xi_1 l_x l_y \Gamma_{xy}^{(z)}(\mathbf{f}) \nonumber \\
 & + & \Xi_2 l_x l_y \Gamma_{zx}^{(z)}(\mathbf{f}) \nonumber \\
 & + & \Xi_3 l_x l_y \Gamma_{yz}^{(x)}(\mathbf{f}) \nonumber \\
 & + & \Xi_4 l_x l_y \Gamma_{yz}^{(z)}(\mathbf{f}) \nonumber \\
 & + & \Xi_5 l_x l_y \Gamma_{zx}^{(y)}(\mathbf{f}) \nonumber \\
 & + & \Xi_6 l_z l_x \Gamma_{xy}^{(y)}(\mathbf{f}) \nonumber \\
 & + & \Xi_7 l_z l_x \Gamma_{zx}^{(z)}(\mathbf{f}) \nonumber \\
 & + & \Xi_8 l_y l_z \Gamma_{xy}^{(x)}(\mathbf{f}) \nonumber \\
 & + & \Xi_9 l_y l_z \Gamma_{yz}^{(z)}(\mathbf{l}) \,,
\end{eqnarray}
where coefficients $\Delta$ and $\Xi$ are materials constants.
When enforced by the external field or near a multicritical
point, the antiferromagnetic mode $\mathbf{l}$ and $\mathbf{f}$ can
co-exist, these mixed terms become operational and will allow
the formation of modulated states composed of the two different modes.
Therefore, we can call these modulated states ``improper textures''
as they are enabled by mixed terms coupling modes of differeent symmetry.

Aditionally, there also exist higher order Lifshitz
invariants that are quartic in either $\mathbf{l}$, $\mathbf{f}$
\begin{eqnarray}
w_{4} & = &
\sum_{\alpha=x,y,z}\,\sum_{\beta=x,y}
(\eta_{\alpha}\,f_{\alpha}^2\,\Gamma_{\beta \, z}^{(\beta)}(\mathbf{f})
  +  \tau_{\alpha}\,l_{\alpha}^2\,\Gamma_{\beta \, z}^{(\beta)}(\mathbf{l})
)   \nonumber \\
 & + & \sigma_1 f_x f_y \Gamma_{yz}^{(x)}(\mathbf{f}) \nonumber \\
 & + & \sigma_2 f_x f_y \Gamma_{zx}^{(y)}(\mathbf{f}) \nonumber \\
 & + & \sigma_3 f_y f_z \Gamma_{zx}^{(y)}(\mathbf{f}) \nonumber \\
 & + & \sigma_4 f_z f_x \Gamma_{yz}^{(x)}(\mathbf{f}) \nonumber \\
 & + & \sigma_5 l_x l_y \Gamma_{yz}^{(x)}(\mathbf{l}) \nonumber \\
 & + & \sigma_6 l_x l_y \Gamma_{zx}^{(y)}(\mathbf{l}) \nonumber \\
 & + & \sigma_7 l_y l_z \Gamma_{zx}^{(y)}(\mathbf{l}) \nonumber \\
 & + & \sigma_8 l_z l_x \Gamma_{yz}^{(x)}(\mathbf{l}) \,.
\end{eqnarray}
Depending on many materials parameters $a_{x,yz},b_{x,y,z},\Delta,\Xi, \eta,\tau$, and $\sigma$,
these terms describe possible coupled modulations of coexisting
primary symmetry modes $\mathbf{l},\mathbf{f}$, which can
take place in distinct fashion in all three spatial directions.
In addition the coupling can display markedly anharmonic effects.

We observe that the Lifshitz-(type)-invariants couple different Cartesian
components of the order-parameters to different spatial directions via the
gradient term. Thus, these terms break isotropy in spin-space.
This implies that their microscopic origin is the relativistic
spin-orbit interaction. Specifically, a microscopic mechanism to explain
these terms is the antisymmetric pairwise Dzyaloshinskii-Moriya exchange,
or appropriate generalization for magnetic systems with
a more itinerant character of spin-ordering,

A complete Landau-Ginzburg free energy density for a polar
metamagnet would collect all these terms

\begin{equation}
w = w_E + w_0 + w_D + w_F + w_a + w_{\mu} + w_{\Delta} + w_4 \,
\end{equation}

Minimizing this free energy functional maps
out all possibilities of metamagnetic modulations
around a tricritical point for the specific antiferromagnetic order parameter in space group Cmc2$_1$.
The corresponding Euler-Lagrange equations for the
variational problem will constitute a system of
coupled partial differential equations for the
degrees of freedom described by the fields $\mathbf{l}(\mathbf{r})$,
and $\mathbf{f}(\mathbf{r})$.

A dedicated theory for a specific material could
be distilled from this most general functional
by restricting to a few crucial terms.
For a simple case, which may pertain to Ca$_3$Ru$_2$O$_7$,
it may suffice to consider
only one-dimensional modulations in $y$-direction and
spin-components in the $yz$-plane. This is the case
sketched in Fig.4b.
The most important terms are then proper Lifshitz invariant
$ F_y\,\Gamma_{yz}^{(y)})(\mathbf{f}) $ and the mixed
Lifshitz-invariants
$a_{z} \, f_{z}^2\,\Gamma_{yz}^{(y)})(\mathbf{l})$
and
$(b_{y} \, l_y^2 + b_z\, l_z^2)\,\Gamma_{y \, z}^{(y)}(\mathbf{f})$.
These inhomogeneous contributions to the
free energy imply
that the presence of a net magnetization $f_z$
in an applied field along the polar axis favours an instability
towards an antiferromagnetic modulation in the $yz$-plane.
But, the local ferromagnetic modulation
is also unstable with respect to modulations through
the proper Lifshitz invariants. Only
strong anisotropies can prevent the instability
of the spin-system towards mixed states
where ferromagnetic and antiferromagnetic
configurations are simultaneously present in
a spatially modulated fashion.
The presence of these different effective couplings
then leads to modulations with a competing character,
as different coupling terms co-operate and frustrate each
other. In our observations, this competing character of
the modulation is noticable, as the characteristic
modulation length displays a pronounced temperature dependence.
For an ordinary Dzyaloshinskii spiral, this behavior is
unusual and unexpected\cite{Dzyaloshinskii1964}, as in
that case there is only one coupling term that rules
the frustration of one simple symmetry mode.
Also, the presence of the higher order term
could lead to marked anharmonicities in metamagnetic textures that
are driven by the higher order mixed terms.

For the particular antiferromagnetic order
in Ca$_3$Ru$_2$O$_7$, the mixed Lifshitz-type
terms are of higher order and affect the magnetic
spin-structure only in the region of a metamagnetic
co-existence. It is the underlying tricritical point
which reveals the presence of these terms, Fig.4c.
However, appropriate symmetry of an antiferromagnetic
mode can also allow mixed Lifshitz type invariants
with a bilinear form like,
\begin{equation}
C_{ij}^{(\gamma)}\,(f_i\, \partial_{\gamma} l_j - l_j \partial_{\gamma} f_i )
.
\end{equation}
In the vicinity of the bi-critical point,
in the case of a system with weak anisotropy, sketched in Fig.~4~d,
these terms drive the existence of metamagnetic textures
with modulation between antiferromagnetic ground-state
in spin-flopped configuration and the field-enforced ferromagnetism.